\documentclass[aip,graphicx]{revtex4-1}

\draft 

\usepackage{amsmath}
\usepackage{amssymb}

\begin{document}


\title{Anisotropic turbulent viscosity and large-scale motive
force \\
in thermally driven turbulence at low Prandtl number} 



\author{K.A. Mizerski}
\email[]{kamiz@igf.edu.pl}
\affiliation{Department of Magnetism, Institute of Geophysics, Polish Academy of Sciences, Ksiecia Janusza 64, 01-452, Warsaw, Poland}


\date{\today}

\begin{abstract}
The fully developed turbulent Boussinesq convection is known to form large-scale rolls, often termed the 'large-scale circulation' (LSC). It is an interesting question how such a large-scale flow is created,
in particular in systems when the energy input occurs at small scales,
when inverse cascade is required in order to transfer energy into
the large-scale modes. Here, the small-scale driving is introduced
through stochastic, randomly distributed heat source (say radiational).
The mean flow equations are derived by means of simplified renormalization
group technique, which can be termed 'weakly nonlinear renormalization
procedure' based on consideration of only the leading order terms
at each step of the recursion procedure, as full renormalization in
the studied anisotropic case turns out unattainable. The effective,
anisotropic viscosity is obtained and it is shown, that the inverse
energy cascade occurs via an effective 'motive force' which takes
the form of transient negative, vertical diffusion.
\end{abstract}


\maketitle 

\section{Introduction}
The investigation of turbulent flows involves description of a very complicated nonlinear dynamics of small scale fluctuations, hence it is extremely difficult and requires sophisticated mathematical tools. In particular the emergence of large-scale coherent structures is a topic of interest, i.e. the transfer of energy from small scales to the large scales termed the inverse energy cascade. To simplify the problem various assumptions have been put forward and
in particular a common simplification in the theoretical approaches is the assumption of the so-called weak turbulence, which corresponds to weak amplitude of turbulent pulsations and linearization of their evolution, cf. \cite{Zakharov_ea_1992, NewellRumpf2011, Nazarenko2011}. Such an approach lacks the crucial effect of nonlinear dynamics
of the fluctuations. As argued in \cite{Monsalve_ea_2020} or \cite{Tobias_ea_2013} in some cases the regime of weak turbulence can be sustained for long times, nevertheless, it is much more common for natural systems to develop into the strong turbulence regime, where the evolution of turbulent fluctuations becomes non-linear. In order to treat the
fully nonlinear regime and reliably estimate turbulent diffusion the renormalization technique has been developed and applied to strongly turbulent flows. This is a statistical closure approximation which is based on systematic, subsequent (iterative) elimination of thin
wave-number bands from the Fourier spectrum of rapidly evolving variables (cf. \cite{Wyld1961, MaMazenko1975, Forster_ea_1977}). Notable contributions come from \cite{YO1986, SW1998, McComb2014} who have published comprehensive works on renormalization of hydrodynamic equations. 

Another powerful method which allows to relate the turbulent diffusion to the turbulent energy tensor is the so-called two-scale direct-interaction approximation (TSIDA) dating back to Kraichnan (1959, 1965) \cite{Kraichnan1959, Kraichnan1965}. It involves introduction
of a tensorial response function to an infinitesimal impulse-force and application of a two-scale approach in space and time related by the same parameter of expansion. Despite its limitations it allows to describe the turbulent viscosity in strong turbulence once the statistical properties of the underlying small-scale chaotic flow
are known, see \cite{Yoshizawa1998, Yokoi2020} for a review. 

Recent investigations of \cite{Mizerski2020, Mizerski2021a} involved applications of the renormalization group method to study the effect of non-stationarity and anisotropy on the magnetohydrodynamic turbulence in what could be called an 'intermediate regime of turbulence' or 'weakly nonlinear
turbulence', in contrast to simple, linear, weak turbulence regime. Due to high complexity of the mathematical approach in the case of non-stationary and non-isotropic turbulence the effect of nonlinear evolution of the fluctuations has been included only at leading order at each step of the renormalization procedure. As a result, although reliable estimates of the turbulent electromotive force could be made
the wave number dependence of all the turbulent coefficients likewise of the energy and helicity spectra was not fully resolved. Because the problem of turbulent convection is also anisotropic due to action of vertical buoyancy, full renoramalization of the Boussinesq equations turns out unattainable thus the latter approach corresponding to 'intermediate turbulence' is adopted here in order to study the physical mechanism of formation of the large-scale convection cells from small-scale energy input. The nonlinear evolution of turbulent pulsations is thus included through calculation of leading order expressions for the effective turbulent viscosity at each step of the iterative renormalization procedure,
which corresponds to expansion of the full, renormalized coefficients up to first order in the Reynolds number. The turbulence is assumed to be driven by a small-scale, statistically random (Gaussian) heat source and the Prandtl number, that is the ratio of the viscosity to thermal diffusivity is assumed small, so that the temperature dynamics is dominated by rapid diffusion and the heat source. It is worth mentioning, that random heat sources are considered e.g. in the dynamics of dusty media as stochastically heated dust grains play an important role
in transport of radiation (cf. \cite{Camps_ea_2015}) and in the problem of stochastic heat engines (cf. \cite{Serra-Garcia_ea_2016}). Renormalized dynamical equations for the mean flow are obtained which contain turbulent coefficients describing the net nonlinear effect of short-wavelength fluctuations,
such as the turbulent viscosity and the turbulent coefficient describing the effective 'motive force' at large scales, which takes the form of negative vertical diffusion in the studied regime. The results correspond to a somewhat initial stage of formation of the large-scale convective flow, as the 'intermediate turbulence' regime is necessarily eventually destroyed and strong turbulence must emerge.

The dynamics of the turbulent Boussinesq convection involves formation of large-scale circulation (LSC) or the so-called 'wind of turbulence' (cf. \cite{GrossmannLohse2000, FunfschillingAhlers2004, BrownAhlers2009, Xi_ea_2009, Horn_ea_2022}). In cylindrical geometry with comparable vertical and horizontal size, the LSC is believed
to result from a quasi-two-dimensional, coupled horizontal sloshing and torsional (ST) oscillatory mode. Roche (2020) \cite{Roche2020} studied the physics of transition to the 'ultimate state' of convection at very high Rayleigh number (a measure of relative strength of the buoyancy force with
respect to diffusion) and developed a model based on boundary layer stability. Zwirner et al. (2020) \cite{Zwirner_ea_2020} suggested, that such transitions could occur through the development of elliptical instabilities and showed that states with smaller amount of large-scale rolls built on top of each other transport heat more efficiently than states with more complex roll-structure. Vasiliev et al. (2019) \cite{Vasiliev_ea_2019} discovered for the first time spontaneous formation of large-scale azimuthal flow. Here, we analyse the system driven thermally by a random heat source and in that way we avoid the problem of boundary conditions and thus the effect of boundary layers; we do not study the LSC structure, but simply study the physical mechanism of LSC formation, i.e. derive
formula for the effective 'motive force' driving LSC, which is shown to take the form of negative vertical diffusion. This exact form is valid only as long as the regime of weak and 'intermediate' turbulence persists, that is at the initial and intermediate stages of evolution
and turbulence development, but once the turbulence becomes strong the structure of the 'motive' force is expected to change. This is a fundamental study and the result sheds light on the physics of the process of energy transfer to large scales in thermal convection. The introduced simplification can be viewed as an advantage in the
sense, that the problem of inverse turbulent energy cascade in convection is extracted and studied in isolation from the influence of velocity boundary conditions.

\section{Dynamical equations and mathematical formulation}

To study the thermally driven turbulence in an incompressible fluid we consider a fluid layer between two flat, parallel boundaries distant $L$ apart, with gravity $\mathbf{g}=-g\hat{\mathbf{e}}_{z}$ pointing
downwards and volume heat sources delivering heat to the system at a rate $\mathcal{Q}(\mathbf{x},t)$. Such a system is governed by the following dynamical equations describing the evolution of the velocity field of the fluid flow $\mathbf{u}(\mathbf{x},t)$ and the temperature field $T(\mathbf{x}.t)$ under the Boussinesq approximation
(cf. \cite{SpiegelVeronis1960, Mizerski2021b})
\begin{subequations}
\begin{equation}
\frac{\partial\mathbf{u}}{\partial t}+\left(\mathbf{u}\cdot\nabla\right)\mathbf{u}=-\nabla\Pi+g\bar{\alpha}T\hat{\mathbf{e}}_{z}+\nu\nabla^{2}\mathbf{u},\label{eq:NS}
\end{equation}
\begin{equation}
\frac{\partial T}{\partial t}+\mathbf{u}\cdot\nabla T=\kappa\nabla^{2}T+Q,\label{eq:IND}
\end{equation}
\begin{equation}
\nabla\cdot\mathbf{u}=0,\label{eq:DIVS}
\end{equation}
\end{subequations}
where $\Pi=p/\bar{\rho}$ is the pressure divided
by the mean density $\bar{\rho}$, $\kappa=k/\bar{\rho}\bar{c}_{p}=\mathrm{const.}$
is the thermal diffusivity of the fluid, uniform by assumption ($k$ is the fluid's thermal conductivity) and $\bar{T}$ is the mean temperature of the system which within the Boussinesq approximation is much greater
than any temperature variations; $\bar{c}_{p}$, $\bar{\alpha}$ are the mean values of the specific heat at constant pressure and the coefficient of thermal expansion, respectively. The kinematic viscosity of the fluid is denoted by $\nu$ and $Q=\mathcal{Q}/\bar{\rho}\bar{c}_{p}$ is the heat source in $K/s$. The solenoidal constraint for the velocity field (\ref{eq:DIVS}) simply expresses the law of mass conservation.
As typically done in the case of Boussinesq convection we have also assumed, that the adiabatic temperature gradient $g\bar{\alpha}\bar{T}/\bar{c}_{p}\ll\left\Vert \nabla T\right\Vert $ is much smaller than the typical temperature gradients in the fluid flow (typically about a thousand times smaller in laboratory flows, cf. e.g. \cite{Mizerski2021b}). 

We concentrate on fluids with low Prandtl numbers
\begin{equation}
Pr=\frac{\nu}{\kappa}\ll1,\label{eq:Pr_small}
\end{equation}
such as e.g. liquid gallium and write down the equations in the following
non-dimensional form\begin{subequations}
\begin{equation}
\frac{\partial\mathbf{u}}{\partial t}+\epsilon\left(\mathbf{u}\cdot\nabla\right)\mathbf{u}=-\nabla\Pi+T\hat{\mathbf{e}}_{z}+\nabla^{2}\mathbf{u},\label{eq:NS-3}
\end{equation}
\begin{equation}
Pr\frac{\partial T}{\partial t}+Pe\mathbf{u}\cdot\nabla T=\nabla^{2}T+Q,\label{eq:IND-3}
\end{equation}
\begin{equation}
\nabla\cdot\mathbf{u}=0,\label{eq:DIVS-3}
\end{equation}
\end{subequations}where the Reynolds number $(Re=)\epsilon$ and
the P$\acute{\textrm{e}}$clet number $Pe$ are defined in a standard
form
\begin{equation}\label{RePePr}
\epsilon=\frac{UL}{\nu},\qquad Pe=\epsilon Pr=\frac{UL}{\kappa},
\end{equation}
and we have chosen $L^{2}/\nu$ for the time scale, $L$ for the spatial
scale, $\nu U/g\bar{\alpha}L^{2}$ for the temperature scale, $\kappa\nu U/g\bar{\alpha}L^{4}$
for the heat source scale and finally pressure was scaled with $\bar{\rho}\nu U/L$.
We will seek for the form of the large-scale flow equations in the
limit of 'intermediate turbulence' (described in general terms in
the introduction and in detail below and in \cite{Mizerski2020, Mizerski2021a}),
through expansions in the Reynolds number $\epsilon$; since $Pr\ll1$
it follows, that the P$\acute{\textrm{e}}$clet number must also be
small, $Pe\ll1$. Hence the final set of equations describing convection
at low Prandtl number takes the form\begin{subequations}
\begin{equation}
\frac{\partial\mathbf{u}}{\partial t}+\epsilon\left(\mathbf{u}\cdot\nabla\right)\mathbf{u}=-\nabla\Pi+T\hat{\mathbf{e}}_{z}+\nabla^{2}\mathbf{u},\label{eq:NS-3-1}
\end{equation}
\begin{equation}
\nabla^{2}T=-Q,\qquad\nabla\cdot\mathbf{u}=0.\label{eq:IND-3-1}
\end{equation}
\end{subequations}Furthermore, we introduce the Fourier transforms
defined in the following way
\begin{subequations}
\begin{equation}
u_{i}(\mathbf{x},t)=\int_{0}^{\Lambda}d^{3}k\int_{-\infty}^{\infty}d\omega\hat{u}_{i}(\mathbf{k},\omega)\mathrm{e}^{\mathrm{i}(\mathbf{k}\cdot\mathbf{x}-\omega t)},\label{eq:F_transform_u}
\end{equation}
\begin{equation}
T(\mathbf{x},t),=\int_{0}^{\Lambda}d^{3}k\int_{-\infty}^{\infty}d\omega\hat{T}(\mathbf{k},\omega)\mathrm{e}^{\mathrm{i}(\mathbf{k}\cdot\mathbf{x}-\omega t)},\label{eq:F_transform_b}
\end{equation}
\begin{equation}
\Pi(\mathbf{x},t)=\int_{0}^{\Lambda}d^{3}k\int_{-\infty}^{\infty}d\omega\hat{\Pi}(\mathbf{k},\omega)\mathrm{e}^{\mathrm{i}(\mathbf{k}\cdot\mathbf{x}-\omega t)},\label{eq:F_transform_U0}
\end{equation}
\begin{equation}
Q(\mathbf{x},t)=\int_{0}^{\Lambda}d^{3}k\int_{-\infty}^{\infty}d\omega\hat{Q}(\mathbf{k},\omega)\mathrm{e}^{\mathrm{i}(\mathbf{k}\cdot\mathbf{x}-\omega t)},\label{eq:F_transform_B0}
\end{equation}
\end{subequations}
where according to standard renormalization approach
we have introduced the upper cut-off for the Fourier spectra $\Lambda$,
which in natural systems appears due to enhanced energy dissipation
at very small scales of turbulence.

The aim of this analysis is to study the large-scale flows in turbulent
convection at low $Pr$. In order to model developed turbulence we
assume a stochastic heat source, Gaussian with zero mean
\begin{equation}
\left\langle Q\right\rangle =0,\label{eq:zero_mean_HS}
\end{equation}
statistically homogeneous and stationary, fully defined by the following
correlation function
\begin{equation}
\left\langle \hat{Q}(\mathbf{k},\omega)\hat{Q}(\mathbf{k}',\omega')\right\rangle =\Xi\left(k\right)\delta(\mathbf{k}+\mathbf{k}')\delta(\omega+\omega'),\label{eq:force_correlations}
\end{equation}
where the function $\Xi\left(k\right)$ will be specified later and
angular brackets denote the ensemble mean, 
\[
\left\langle \cdot\right\rangle -\textrm{ensemble mean}.
\]
Note, that so-induced turbulence will be \emph{anisotropic} because
of action of vertical gravity (buoyancy force). We can calculate a
positive definite quantity
\begin{equation}
\int k^{2}\mathrm{d}\mathring{\varOmega}_{\mathbf{k}}\int\mathrm{d}^{4}q'\left\langle \hat{Q}(\mathbf{k},\omega)\hat{Q}(\mathbf{k}',\omega')\right\rangle =4\pi k^{2}\Xi\left(k\right)>0,\label{eq:F2}
\end{equation}
where $\mathring{\varOmega}_{\mathbf{k}}$ denotes a solid angle associated
with the vector $\mathbf{k}$, which implies $\Xi\left(k\right)>0$.

The approach will be based on the renormalization group technique,
which is an iterative procedure of successive elimination of thin
wave-number bands from the Fourier spectrum of fluctuating fields.
In this way the effect of thin bands of modes with shortest wavelengths
on the remaining modes is calculated at each step of the procedure.
The final aim of this approach is to obtain recursion equations for
coefficients describing the effective mean Reynolds stress $\left\langle \mathbf{u}\mathbf{u}\right\rangle $
as a function of the wave number at each step of the procedure. The
Reynolds stresses are responsible for creation of the turbulent viscosity
and what can be called a 'motive force' for the large-scale flows.
The recursion equations (provided in (\ref{eq:xi-2-1-1}-c)) are then
solved for $k\rightarrow0$ in order to obtain the final forms of
the large-scale viscosity and the motive force which appear in the
mean-field equations and include the effect of nonlinear evolution
of turbulent fluctuations. 

\section{The iterative weakly nonlinear renormalization procedure\label{sec:Renormalization-procedure}}

Introducing new shorter four-component vector notation
\begin{equation}
\mathbf{q}=(\mathbf{k},\omega),\qquad\int_{0}^{\Lambda}d^{3}k\int_{-\infty}^{\infty}d\omega(\cdot)=\int d^{4}q(\cdot),\label{eq:short_notation}
\end{equation}
so that e.g.
\begin{equation}
u_{i}(\mathbf{x},t)=\int d^{4}q\hat{u}_{i}(\mathbf{q})\mathrm{e}^{\mathrm{i}(\mathbf{k}\cdot\mathbf{x}-\omega t)},\label{eq:short_notation_1}
\end{equation}
the equations take the form
\begin{subequations}
\begin{equation}
\left(-\mathrm{i}\omega+k^{2}\right)\hat{u}_{i}(\mathbf{q})-\hat{T}(\mathbf{q})\delta_{i3}+\mathrm{i}k_{i}\hat{\Pi}=-\mathrm{i}\epsilon k_{j}\mathbb{I}_{ij}^{(u)}\left(\mathbf{q}\right)\label{eq:u_Fourier}
\end{equation}
\begin{equation}
k^{2}\hat{T}(\mathbf{q})=\hat{Q}\left(\mathbf{q}\right),\qquad\mathbf{k}\cdot\hat{\mathbf{u}}(\mathbf{q})=0,\label{eq:T_Fourier}
\end{equation}
\end{subequations}
In the above we have also defined the following
convolution integral
\begin{equation}
\mathbb{I}_{ij}^{(u)}=\int d^{4}q'\hat{u}_{i}(\mathbf{q}-\mathbf{q}')\hat{u}_{j}(\mathbf{q}'),\label{eq:Integrals_def}
\end{equation}
which possesses the symmetry property
\begin{equation}
\mathbb{I}_{ij}^{(u)}=\mathbb{I}_{ji}^{(u)}.\label{eq:integral_symmetries}
\end{equation}
It should be noted at this stage, that the convolution integrals,
which represent the nonlinear interactions between fluctuating turbulent
fields are \emph{not} neglected in the evolutional equations for fluctuations
(\ref{eq:u_Fourier},b), and their effect will be included within
the scope of the 'intermediate turbulence regime', based on the iterative
renormalization procedure introduced in \cite{YO1986}.
Thus we go beyond the weak turbulence regime and quantitatively express
the effect of this nonlinearity on the dynamics of the large-scale
flow. However, contrary to the standard renormalization approach at each step
of the iterative procedure based on a step-by-step elimination of
thin wave number bands from the Fourier spectrum, only the terms of
the leading order in $\epsilon$ will be retained; in such a way the
Taylor series in $\epsilon$ will not be finally contracted as in
the renormalization approaches, but instead we will obtain a weakly
nonlinear expressions on effective coefficients appearing in the final
'renormalized' equation for the large-scale flow, i.e. up to the first
order in $\epsilon$.

In order to eliminate pressure we apply the projection operator 
\begin{equation}
P_{ij}(\mathbf{k})=\delta_{ij}-\frac{k_{i}k_{j}}{k^{2}},\label{eq:projection_operator}
\end{equation}
to both sides of the Navier-Stokes equation (\ref{eq:u_Fourier})
\begin{subequations}
\begin{equation}
\gamma\hat{u}_{i}(\mathbf{q})=\frac{P_{i3}}{k^{2}}\hat{Q}(\mathbf{q})-\frac{1}{2}\mathrm{i}\epsilon P_{imn}\mathbb{I}_{mn}^{(u)}\left(\mathbf{q}\right)\label{eq:PijNS}
\end{equation}
\begin{equation}
\mathbf{k}\cdot\hat{\mathbf{u}}(\mathbf{q})=0,\label{eq:kcdot-1}
\end{equation}
\end{subequations}
where we have already introduced $\hat{T}(\mathbf{q})=\hat{Q}(\mathbf{q})/k^{2}$
from (\ref{eq:T_Fourier}) and
\begin{equation}
P_{imn}(\mathbf{k})=k_{m}P_{in}(\mathbf{k})+k_{n}P_{im}(\mathbf{k}),\label{Pimn_def}
\end{equation}
\begin{equation}
\gamma=-\mathrm{i}\omega+k^{2}.\label{eq:gammas_def}
\end{equation}
The smallness of the amplitude of turbulence $\epsilon\ll1$ allows
for proper mathematical formulation of the problem, since the velocity
field is expressed at leading order by the driving $\hat{Q}$ and
the nonlinearity, which is of the order $\mathcal{O}(\epsilon)$,
can be treated in perturbational sense. The iterational procedure
of renormalization is then applicable. The final recursion differential
relations for the coefficients describing the Reynolds stresses can
be solved analytically, thus in particular the turbulent viscosity
and the motive force for the large-scale flow can be determined.

We now start the iterative procedure of taking successive little bites
off the Fourier spectrum from the short-wavelength side in order to
obtain the final nonlinear effect of the fluctuations on the means.
At the first step of the procedure we introduce a parameter $\lambda_{1}$,
which satisfies
\begin{equation}
\delta\lambda=\Lambda-\lambda_{1}\ll1,\label{eq:lambda_bite}
\end{equation}
and divide the Fourier spectrum into two parts
\begin{subequations}
\begin{equation}
\hat{u}_{i}^{>}(\mathbf{k},\omega)=\theta(k-\lambda_{1})\hat{u}_{i}(\mathbf{k},\omega),\quad\textrm{or}\quad\hat{u}_{i}^{>}(\mathbf{k},\omega)=\hat{u}_{i}(\mathbf{k}^{>},\omega),\quad\lambda_{1}<\left|\mathbf{k}^{>}\right|<\Lambda,\label{eq:Fourier_division_1}
\end{equation}
\begin{equation}
\hat{u}_{i}^{<}(\mathbf{k},\omega)=\theta(\lambda_{1}-k)\hat{u}_{i}(\mathbf{k},\omega),\quad\textrm{or}\quad\hat{u}_{i}^{<}(\mathbf{k},\omega)=\hat{u}_{i}(\mathbf{k}^{<},\omega),\quad\left|\mathbf{k}^{<}\right|<\lambda_{1},\label{eq:Fourier_division_2}
\end{equation}
\end{subequations}
and same way for $\hat{Q}$. The equation for the field $\hat{u}_{i}^{<}(\mathbf{k},\omega)$
is obtained by averaging (\ref{eq:PijNS}) over the first shell ($\lambda_{1}<k<\Lambda$)
\begin{align}
\left(-\mathrm{i}\omega+k^{2}\right)\hat{u}_{i}^{<}(\mathbf{q})= & \frac{P_{i3}}{k^{2}}\hat{Q}^{<}(\mathbf{q})-\frac{1}{2}\mathrm{i}\epsilon P_{imn}\mathbb{I}_{mn}^{(u^{<})}\left(\mathbf{q}\right)\nonumber \\
 & -\frac{1}{2}\mathrm{i}\epsilon P_{imn}(\mathbf{k})\int d^{4}q'\left\langle \hat{u}_{m}^{>}(\mathbf{q}')\hat{u}_{n}^{>}(\mathbf{q}-\mathbf{q}')\right\rangle _{c}.\label{eq:u<}
\end{align}
To get the equation for $\hat{u}_{i}^{>}(\mathbf{k},\omega)$ we utilize
(\ref{eq:PijNS}) again 
\begin{align}
\hat{u}_{i}^{>}(\mathbf{q})= & \frac{P_{i3}}{k^{2}\gamma}\hat{Q}^{>}\left(\mathbf{q}\right)-\mathrm{i}\epsilon\frac{P_{imn}}{\gamma}\mathbb{J}_{mn}^{(u)}\left(\mathbf{q}\right)\nonumber \\
 & -\frac{1}{2}\mathrm{i}\epsilon\frac{P_{imn}}{\gamma}\mathbb{I}_{mn}^{(u^{<})}\left(\mathbf{q}\right)+R_{i},\label{eq:u_gtr}
\end{align}
where $\left\langle \cdot\right\rangle _{c}$ denotes conditional
average over the first shell ($\lambda_{1}<k<\Lambda$) statistical
subensemble, described at the beginning of the Appendix (cf. \cite{McComb_ea_1992, McCombWatt1990, McCombWatt1992}); furthermore, we have defined
\begin{equation}
\mathbb{J}_{mn}^{(u)}\left(\mathbf{q}\right)=\int d^{4}q'\hat{u}_{m}^{<}(\mathbf{q}')\hat{u}_{n}^{>}(\mathbf{q}-\mathbf{q}'),\label{eq:Ju}
\end{equation}
and the rest in (\ref{eq:u_gtr}) is given by
\begin{equation}
R_{i}^{(u)}=-\frac{1}{2}\mathrm{i}\epsilon\frac{P_{imn}}{\gamma}\mathbb{I}_{mn}^{\left(u^{>}\right)}.\label{eq:Ru}
\end{equation}
The rests will be neglected on the basis of generating triple order
statistical correlations since they involve only terms second order
in $\mathbf{u}^{>}$ or because of the kept order of accuracy in the
asymptotic limit $\epsilon\ll1$, which will allow to neglect terms
of the order $\mathcal{O}(\epsilon^{3})$; for details the reader
is referred to the Appendix.

In what follows we provide a short description of the asymptotic iterative
procedure, described in detail in the Appendix. First we introduce
(\ref{eq:u_gtr}) into (\ref{eq:u<}) and calculate the dynamical
effect of short-wavelength components $\hat{u}_{i}^{>}(\mathbf{k},\omega)$
on the evolution of $\hat{u}_{i}^{<}(\mathbf{k},\omega)$ (long wavelength
modes). This results in corrections to some of the terms in equation
(\ref{eq:u<}), but also generates terms with a new structure. Therefore
a next step is necessary, involving calculation of the effect of the
next shell $\lambda_{2}=\lambda_{1}-\delta\lambda<k<\lambda_{1}$
(new short-wavelength modes) on the modes with $k<\lambda_{1}-\delta\lambda$
(new long wavelength modes); this is continued until invariance of
the equations for long-wavelength modes is achieved, i.e. the equations
do not change from one iterational step to the next one. We can then
take the limit of infinitesimally narrow wave number bands $\delta\lambda\rightarrow0$,
which leads to differential recursion relations for all the coupling
constants introduced into the equations for long wavelength modes
by couplings of the short wavelength ones. In the isothermal, fully
isotropic case Yakhot and Orszag (1986) \cite{YO1986} calculated the correction
from short wavelength modes in the Navier-Stokes equation which was
proportional to $k^{2}\hat{u}_{i}^{<}\delta\lambda$ thus creating
viscosity correction; the turbulent viscosity was then obtained from
an equation of the form $\mathrm{d}\nu_{turb}/\mathrm{d}\lambda=f(\lambda)$
with an 'initial' condition $\nu_{eff}(\lambda=\Lambda)=\nu$. The
case at hand is \emph{anisotropic} because of the vertical gravity
(buoyancy force); explicit calculation of two initial steps of the
renormalization procedure is enough to derive the final differential
recursion relations with satisfactory accuracy. The details of the
procedure are provided in the Appendix.

\section{Dynamics of the large-scale flow\label{subsec:Dynamics-of-mean}}

The mean-field equations are derived in the Appendix A and take the
following form
\begin{subequations}
\begin{equation}
\frac{\partial\left\langle \mathbf{U}\right\rangle }{\partial t}+\left(\left\langle \mathbf{U}\right\rangle \cdot\nabla\right)\left\langle \mathbf{U}\right\rangle =-\nabla\left\langle \Pi\right\rangle -9\varsigma\nabla^{2}\left\langle U\right\rangle _{z}\hat{\mathbf{e}}_{z}+\left(\nu+6\varsigma\right)\nabla_{h}^{2}\left\langle \mathbf{U}\right\rangle +\left(\nu+4\varsigma\right)\frac{\partial^{2}\left\langle \mathbf{U}\right\rangle }{\partial z^{2}}\label{eq:MF_eq_WN-2}
\end{equation}
\begin{equation}
\nabla\cdot\left\langle \mathbf{U}\right\rangle =0,\label{eq:divMF}
\end{equation}
\end{subequations}
where the coefficient
\begin{equation}
\varsigma=\frac{2\pi^{2}}{735}\frac{g^{2}\bar{\alpha}^{2}L^{3}\mathbb{Q}^{2}}{\nu^{3}\kappa^{2}k_{\ell}^{7}},\label{eq:varsigma-1}
\end{equation}
includes the effect of the turbulent fluctuations on the means; in
the above $\mathbb{Q}$ is the magnitude of heat delivery rate (in
$K/s$) and $k_{\ell}$ is the wave number based on the length scale
of most energetic turbulent eddies. The general differential recursion
relations for the turbulent coefficients are solved in the Appendix
A, see (\ref{eq:xi-2-1-1}-c). It is evident, that the term $-9\varsigma\nabla^{2}\left\langle U\right\rangle _{z}\hat{\mathbf{e}}_{z}$
is the large-scale motive force responsible for energy transfer from
small scales to large scales, i.e. the inverse energy cascade; it
takes the form of negative diffusion in the vertical direction, which
drives the large-scale flow. 

For the sake of a rough estimate we may take the Kolmogorov cut-off
value $\Lambda\sim(\sqrt{gL}/\nu)^{3/4}L^{-1/4}$, where the free-fall
velocity $\sqrt{gL}$ was used as the convective velocity scale, which
yields $\varsigma\sim\nu\mathcal{G}^{-13/4}\mathcal{H}^{2}\left(k_{\ell}/\Lambda\right)^{-7}$,
with $\mathcal{H}=g^{1/2}\bar{\alpha}L^{7/2}\mathbb{Q}/\kappa\nu$
and $\mathcal{G}=\sqrt{gL^{3}/\nu^{2}}$. Since $k_{\ell}/\Lambda\ll1$,
this coefficient is expected to be much larger than the molecular
viscosity, in particular in strongly driven turbulence $\mathcal{H}\gg1$.
It is, in fact a typical situation when the turbulent coefficients
greatly exceed the molecular ones.

\subsection{Linear regime}

Linearization of (\ref{eq:MF_eq_WN-2}) and substitution of normal
modes in the form
\begin{subequations}
\begin{equation}
\left\langle U\right\rangle _{z}=\mathrm{e}^{\sigma t}\cos\left(\mathbf{K}_{h}\cdot\mathbf{x}\right)\sin\left(K_{z}z\right)\hat{U}_{z},\label{eq:Fourier_lin}
\end{equation}
\begin{equation}
\left\langle \mathbf{U}\right\rangle _{h}=\mathrm{e}^{\sigma t}\sin\left(\mathbf{K}_{h}\cdot\mathbf{x}\right)\cos\left(K_{z}z\right)\hat{\mathbf{U}},\label{eq:Fourier_lin_h}
\end{equation}
\begin{equation}
\mathbf{K}\cdot\hat{\mathbf{U}}=0,\label{eq:Fourier_lin_div}
\end{equation}
\begin{equation}
\left\langle \Pi\right\rangle =\mathrm{e}^{\sigma t}\cos\left(\mathbf{K}_{h}\cdot\mathbf{x}\right)\cos\left(K_{z}z\right)\hat{\mathbf{\Pi}},\label{eq:Fourier_lin_p}
\end{equation}
\end{subequations}
leads to
\begin{equation}
\sigma\left\langle \mathbf{U}\right\rangle =-\mathrm{i}\mathbf{K}\hat{\Pi}-\nabla\left\langle \Pi\right\rangle +9\varsigma K^{2}\left\langle U\right\rangle _{z}\hat{\mathbf{e}}_{z}-\left(\nu+6\varsigma\right)K_{h}^{2}\left\langle \mathbf{U}\right\rangle -\left(\nu+4\varsigma\right)K_{z}^{2}\left\langle \mathbf{U}\right\rangle ,\label{eq:MF_eq_WN-1-2}
\end{equation}
which under action of the projection operator $\mathbf{1}-\mathbf{K}\mathbf{K}/K^{2}$
where $\mathbf{1}$ is the unity matrix transforms into
\begin{equation}
\sigma\left\langle U\right\rangle _{i}=9\varsigma K^{2}\left(\delta_{i3}-\frac{K_{i}K_{z}}{K^{2}}\right)\left\langle U\right\rangle _{z}-\left(\nu+6\varsigma\right)K_{h}^{2}\left\langle U\right\rangle _{i}-\left(\nu+4\varsigma\right)K_{z}^{2}\left\langle U\right\rangle _{i},\label{eq:MF_eq_WN-1-2-1}
\end{equation}
so that the pressure is eliminated. This yields for the growth rate
of the mean flow
\begin{equation}
\sigma=\left(3\varsigma-\nu\right)K_{h}^{2}-\left(\nu+4\varsigma\right)K_{z}^{2}.\label{eq:g_rate}
\end{equation}
It follows, that in turbulent convection driven by strong stochastic
heat sources $\mathcal{H}\gg1$ the growth rate takes the approximate
form
\begin{equation}
\sigma\approx3\varsigma K_{h}^{2}-4\varsigma K_{z}^{2},\label{eq:grate_approx}
\end{equation}
and thus turbulence excites large-scale modes with horizontal wavelengths
shorter than vertical ones,
\begin{equation}
\sigma>0\;\Leftrightarrow\;K_{z}^{2}<\frac{3}{4}K_{h}^{2}\;\Leftrightarrow\;L_{h}<\frac{\sqrt{3}}{2}L_{z}\approx0.87L_{z}.\label{eq:sigma_Ls}
\end{equation}
In other words in the studied problem the large scale flow is expected
to form vertically elongated rolls. The growth rate increases unboundedly
with $K_{h}^{2}$, but since the equation (\ref{eq:MF_eq_WN-2}) describes
the large-scale flow only, there is a natural upper bound on the horizontal
wave number of the large-scale modes and thus on the growth rate.
As argued in the Appendix B there also exists an additional term on
the r.h.s. of the mean flow equation (\ref{eq:MF_eq_WN-2}) of the
form $\varrho\partial_{z}^{2}\left\langle U\right\rangle _{z}\hat{\mathbf{e}}_{z}$,
which is of smaller (asymptotically negligible) magnitude than the
other turbulent terms proportional to the coefficient $\varsigma$,
i.e. $\varrho<\varsigma$. Thus the growth rate is modified to $\sigma=3\varsigma K_{h}^{2}-4\varsigma K_{z}^{2}-\varrho\frac{K_{z}^{2}K_{h}^{2}}{K^{2}}$,
but the sign of the turbulent coefficient $\varrho$ remains undetermined,
hence it is not clear whether it acts as additional diffusion (if
$\varrho>0$) or additional motive force (if $\varrho<0$).

The normal modes in the form (\ref{eq:Fourier_lin}) are individually
also solutions of the nonlinear equation (\ref{eq:MF_eq_WN-2}). Of
course in the problem at hand, the energy is transferred from the
small scale fluctuations, where the flow is thermally driven, to the
large-scale modes and thus in the limit of small $K$ the dynamics
naturally involves wave packets, rather than individual modes, which
evolve nonlinearly. Still it is possible for the most unstable modes
to dominate the dynamics, in which case the amplitude of convection
grows unboundedly in time until the initial assumption of small Reynolds
number ceases to be valid and saturation may occur. In other words
the analysis of weakly nonlinear turbulence does not lead to saturation
of large-scale modes, which is possible only beyond the scope of this
approach, that is in fully developed, strong turbulence.

\section{Conclusions}

The presented analysis was focused on derivation of the effective
equation describing the dynamics of the large-scale flow (circulation)
in turbulent convection driven by a random heat source at low $Pr$.
The applied technique was based on the renormalization approach of
\cite{YO1986} and \cite{McComb_ea_1992} (see also
\cite{SW1998} for a review of the method), which allowed
to incorporate the effect of the nonlinear terms in the dynamical
equations for small-scale turbulent fluctuations, and calculate the
anisotropic turbulent viscosity and 'motive force' induced by the
fluctuations and experienced by the large-scale flows. The renormalized
mean-flow equation was derived and it was shown, that the 'motive
force' acts in the form of negative vertical diffusion, $\partial_{t}\left\langle \mathbf{U}\right\rangle =-9\varsigma\nabla^{2}\left\langle U\right\rangle _{z}\hat{\mathbf{e}}_{z}+\dots$,
where $\varsigma$ is given in (\ref{eq:varsigma-1}), leading to
enhancement of the mean flow energy. The general recursion differential
equations for all the turbulent coefficients are provided in (\ref{eq:xi-2-1}-c)
for any form of the random heat-source function $\Xi(k)$.
\\
\\
{\bf{Data availability statement:}}
All data generated or analysed during this study are included in this published article

\begin{acknowledgments}
The support of the National Science Centre of Poland (grant
no. 2017/26/E/ST3/00554) is gratefully acknowledged. This work was
partially financed as a part of the statutory activity from the subvention
of the Ministry of Science and Higher Education in Poland.
\end{acknowledgments}

\appendix
\section{Details of the iterative, weakly nonlinear renormalization procedure}

The details of the renormalization procedure applied in order to obtain
the mean field equations are given in here. First of all we clarify
how the ensemble averaging should be understood and explain the concept
of a conditional average over a statistical subensemble for short-wavelength
modes. We adopt the approach of McComb \emph{et al}. (1992) \cite{McComb_ea_1992} (cf. also
\cite{McCombWatt1990, McCombWatt1992}). The essential idea of this approach is
the introduction of a subensemble of flow realizations including near-chaotic
statistical properties for the short-wavelength shell $\lambda_{1}<k\leq\Lambda$,
but remaining quasi-deterministic for $k\leq\lambda_{1}$. The subensemble
average can be precisely defined and then, utilizing the assumption,
that in the turbulent cascade the energy transfer in the Fourier space
is local (i.e. the assumption of ergodicity of the system), the following
crucial properties can be proved
\begin{subequations}
\begin{equation}
\left\langle \hat{\mathbf{u}}^{<}\left(\mathbf{q}\right)\right\rangle _{c}=\hat{\mathbf{u}}^{<}\left(\mathbf{q}\right),\qquad\left\langle \hat{\mathbf{u}}^{<}\left(\mathbf{q}\right)\hat{\mathbf{u}}^{<}\left(\mathbf{q}'\right)\right\rangle _{c}\approx\hat{\mathbf{u}}^{<}\left(\mathbf{q}\right)\hat{\mathbf{u}}^{<}\left(\mathbf{q}'\right),\label{eq:subensamble_av_prop_1}
\end{equation}
\begin{equation}
\left\langle \hat{\mathbf{u}}^{>}\left(\mathbf{q}'\right)\right\rangle _{c}\approx\left\langle \hat{\mathbf{u}}^{>}\left(\mathbf{q}'\right)\right\rangle =0,\qquad\left\langle \hat{\mathbf{u}}^{<}\left(\mathbf{q}\right)\hat{\mathbf{u}}^{>}\left(\mathbf{q}'\right)\right\rangle _{c}\approx\hat{\mathbf{u}}^{<}\left(\mathbf{q}\right)\left\langle \hat{\mathbf{u}}^{>}\left(\mathbf{q}'\right)\right\rangle _{c}\approx0,\label{eq:subensamble_av_prop_2}
\end{equation}
\begin{equation}
\left\langle \hat{\mathbf{u}}^{>}\left(\mathbf{q}\right)\hat{\mathbf{u}}^{>}\left(\mathbf{q}'\right)\right\rangle _{c}\approx\left\langle \hat{\mathbf{u}}^{>}\left(\mathbf{q}\right)\hat{\mathbf{u}}^{>}\left(\mathbf{q}'\right)\right\rangle .\label{eq:subensamble_av_prop_3}
\end{equation}
\end{subequations}
For details see particularly section IV and the
beginning of section V in \cite{McComb_ea_1992}.

We now substitute the expressions for short wavelength modes from
(\ref{eq:u_gtr}) into the conditional averages in the equations for
long wavelength modes in (\ref{eq:u<}). Neglecting higher order correlations
of the type $\left\langle \hat{u}_{i}^{>}\hat{u}_{j}^{>}\hat{Q}^{>}\right\rangle _{c}$
etc. (which eliminates the rests in (\ref{eq:u_gtr})) and using $\left\langle \hat{Q}^{>}\right\rangle _{c}=0$
and $\left\langle \hat{u}_{i}^{<}\right\rangle _{c}=\hat{u}_{i}^{<}$
one obtains
\begin{align}
\left\langle \hat{u}_{m}^{>}(\mathbf{q}')\hat{u}_{n}^{>}(\mathbf{q}-\mathbf{q}')\right\rangle _{c}= & \frac{P_{m3}\left(\mathbf{k}'\right)P_{n3}(\mathbf{k}-\mathbf{k}')}{k^{\prime2}\left|\mathbf{k}-\mathbf{k}'\right|^{2}\gamma\left(\mathbf{q}'\right)\gamma(\mathbf{q}-\mathbf{q}')}\left\langle \hat{Q}^{>}\left(\mathbf{q}'\right)\hat{Q}^{>}\left(\mathbf{q}-\mathbf{q}'\right)\right\rangle _{c}\nonumber \\
 & -\frac{\mathrm{i}\epsilon P_{m3}\left(\mathbf{k}'\right)}{k^{\prime2}\gamma(\mathbf{q}')}\frac{P_{npq}(\mathbf{k}-\mathbf{k}')}{\gamma(\mathbf{q}-\mathbf{q}')}\left\langle \hat{Q}^{>}(\mathbf{q}')\mathbb{J}_{pq}^{(u)}\left(\mathbf{q}-\mathbf{q}'\right)\right\rangle _{c}\nonumber \\
 & -\frac{\mathrm{i}\epsilon P_{n3}(\mathbf{k}-\mathbf{k}')}{\left|\mathbf{k}-\mathbf{k}'\right|^{2}\gamma(\mathbf{q}-\mathbf{q}')}\frac{P_{mpq}\left(\mathbf{k}'\right)}{\gamma\left(\mathbf{q}'\right)}\left\langle \hat{Q}^{>}\left(\mathbf{q}-\mathbf{q}'\right)\mathbb{J}_{pq}^{(u)}\left(\mathbf{q}'\right)\right\rangle _{c}\nonumber \\
 & +\mathcal{O}\left(\epsilon^{2}\right)\label{eq:uu_corr}
\end{align}
The first term in (\ref{eq:uu_corr}) is proportional to $\left\langle \hat{Q}^{>}\left(\mathbf{q}'\right)\hat{Q}^{>}\left(\mathbf{q}-\mathbf{q}'\right)\right\rangle _{c}\sim\delta(\mathbf{k})\delta(\omega)$,
hence on taking the inverse Fourier transform of $\frac{1}{2}\mathrm{i}\epsilon P_{imn}(\mathbf{k})\int d^{4}q'\left\langle \hat{u}_{m}^{>}(\mathbf{q}')\hat{u}_{n}^{>}(\mathbf{q}-\mathbf{q}')\right\rangle _{c}$
in the Navier-Stokes equation to return to the real space, this term
vanishes and thus does not contribute to the dynamics of large-scale
fields; it follows, that this term will be disregarded. Substituting
once again for $\hat{\mathbf{u}}^{>}$ from (\ref{eq:u_gtr}) into
the $\mathbb{J}^{(u)}$-terms in (\ref{eq:uu_corr}) and making use
of the symmetry $\mathbf{q}'\mapsto\mathbf{q}-\mathbf{q}'$ under
the integral $\int d^{4}q'$ one obtains
\begin{align}
\int\hspace{-1.5mm}\mathrm{d}^{4}q'\left\langle \hat{u}_{m}^{>}(\mathbf{q}')\hat{u}_{n}^{>}(\mathbf{q}-\mathbf{q}')\right\rangle _{c}= & \,\qquad\qquad\qquad\qquad\qquad\qquad\qquad\qquad\qquad\qquad\qquad\qquad\qquad\qquad\qquad\nonumber \\
 & \hspace{-45mm}\hspace{-5mm}-\mathrm{i}\epsilon\int\hspace{-1.5mm}\mathrm{d}^{4}q'\hspace{-1.5mm}\int\hspace{-1.5mm}\mathrm{d}^{4}q''\frac{\hat{u}_{p}^{<}(\mathbf{q}'')P_{m3}\left(\mathbf{k}'\right)P_{npq}(\mathbf{k}-\mathbf{k}')P_{q3}(\mathbf{k}-\mathbf{k}'-\mathbf{k}'')}{k^{\prime2}\left|\mathbf{k}-\mathbf{k}'-\mathbf{k}''\right|^{2}\gamma(\mathbf{q}')\gamma(\mathbf{q}-\mathbf{q}')\gamma(\mathbf{q}-\mathbf{q}'-\mathbf{q}'')}\left\langle \hat{Q}^{>}(\mathbf{q}')\hat{Q}^{>}\left(\mathbf{q}-\mathbf{q}'-\mathbf{q}''\right)\right\rangle _{c}\nonumber \\
 & \hspace{-45mm}\hspace{-5mm}+(m\leftrightarrow n)+\mathcal{O}\left(\epsilon^{2}\right),\label{eq:uu_corr-1}
\end{align}
where $(m\leftrightarrow n)$ in (\ref{eq:uu_corr-1}) denotes a term
of the same structure as the previous one but with exchanged indices
$m$ and $n$. We can now substitute for the heat-source correlations
\begin{equation}
\left\langle \hat{Q}(\mathbf{k},\omega)\hat{Q}(\mathbf{k}',\omega')\right\rangle =\Xi\left(k\right)\delta(\mathbf{k}+\mathbf{k}')\delta(\omega+\omega'),\label{eq:force_correlations-1}
\end{equation}
cf. (\ref{eq:force_correlations}), into (\ref{eq:uu_corr-1}) and
perform the $\mathbf{q}''$ integral which yields
\begin{align}
\int\hspace{-1.5mm}\mathrm{d}^{4}q'\left\langle \hat{u}_{m}^{>}(\mathbf{q}')\hat{u}_{n}^{>}(\mathbf{q}-\mathbf{q}')\right\rangle _{c}= & -\mathrm{i}\epsilon\hat{u}_{p}^{<}(\mathbf{q})\int\hspace{-1.5mm}\mathrm{d}^{4}q'\frac{\Xi\left(k'\right)P_{m3}\left(\mathbf{k}'\right)P_{npq}(\mathbf{k}-\mathbf{k}')P_{q3}(\mathbf{k}')}{k^{\prime4}\left|\gamma(\mathbf{q}')\right|^{2}\gamma(\mathbf{q}-\mathbf{q}')}\nonumber \\
 & +(m\leftrightarrow n)+\mathcal{O}\left(\epsilon^{2}\right).\label{eq:uu_corr-1-2}
\end{align}
The $q'$-integrals are taken over an intersection of the domains
$\lambda_{1}<k'<\Lambda$ and $\lambda_{1}<\left|\mathbf{k}-\mathbf{k}'\right|<\Lambda$,
i.e. 
\begin{equation}
\left\{ \mathbf{k}':\;\lambda_{1}<k'<\Lambda,\;\lambda_{1}<\left|\mathbf{k}-\mathbf{k}'\right|<\Lambda\right\} .\label{eq:qprime_domain}
\end{equation}
Following the approach of Yakhot and Orszag (1986) \cite{YO1986} and Smith and Woodruff
(1998) \cite{SW1998} we calculate the $q'$-integrals to lowest nontrivial order
in the distant-interaction limit 
\begin{equation}
\frac{k}{k'}\rightarrow0,\qquad\frac{\omega}{\omega'}\rightarrow0,\label{eq:dist_interact_limit}
\end{equation}
which stems from the assumption of local energy transfer in the Fourier
spectrum of a turbulent cascade. The integrals are then calculated
by setting $\omega=0$ and substitution $\mathbf{k}'\mapsto\mathbf{k}'+\mathbf{k}/2$
hence by symmetrization of the integration domain; in the case at
hand, when the zeroth order term $\sim(k/k')^{0}$ vanishes no corrections
of the order $k$ (and higher) from the integration domain are then
necessary, and it simplifies to
\begin{equation}
\left\{ \mathbf{k}':\;\lambda_{1}<k'<\Lambda\right\} .\label{eq:qprime_domain-2}
\end{equation}
This way the total renormalized corrections from short-wavelength
modes in are proportional to $k^{2}$, which implies that the lowest
non-trivial order in distant interactions produces corrections to
diffusivities.

Therefore the corrections from short-wavelength modes to the equations
for long wavelength fluctuations in (\ref{eq:uu_corr-1-2}) can be
expressed as follows. Making the aforementioned substitution $\mathbf{k}'\rightarrow\mathbf{k}'+\frac{1}{2}\mathbf{k}$
to symmetrize the domain of integration one obtains
\begin{align}
\int\mathrm{d}^{4}q'\left\langle \hat{u}_{m}^{>}(\mathbf{q}')\hat{u}_{n}^{>}(\mathbf{q}-\mathbf{q}')\right\rangle _{c}= & \,\qquad\qquad\qquad\qquad\qquad\qquad\qquad\qquad\qquad\qquad\qquad\qquad\qquad\qquad\qquad\nonumber \\
 & \hspace{-40mm}\hspace{-5mm}-\mathrm{i}\epsilon\hat{u}_{p}^{<}(\mathbf{q})\int_{\lambda_{1}}^{\Lambda}\hspace{-1.5mm}\mathrm{d}k'\hspace{-1.5mm}\int\hspace{-1.5mm}\mathrm{d}\mathring{\Omega}\hspace{-1.5mm}\int_{-\infty}^{\infty}\hspace{-2mm}\mathrm{d}\omega'\frac{\Xi\left(\left|\mathbf{k}'+\frac{1}{2}\mathbf{k}\right|\right)P_{m3}\left(\mathbf{k}'+\frac{1}{2}\mathbf{k}\right)P_{npq}(\frac{1}{2}\mathbf{k}-\mathbf{k}')P_{q3}(\mathbf{k}'+\frac{1}{2}\mathbf{k})}{\left(k^{\prime2}+k_{r}^{\prime}k_{r}\right)\left|\gamma(\mathbf{k}'+\frac{1}{2}\mathbf{k},\omega')\right|^{2}\gamma(\mathbf{k}'-\frac{1}{2}\mathbf{k},-\omega')}\nonumber \\
 & \hspace{-40mm}\hspace{-5mm}+(m\leftrightarrow n)+\mathcal{O}\left(\epsilon^{2}\right),\label{eq:uu_corr-1-2-1}
\end{align}
where $\mathring{\varOmega}$ denotes a solid angle. 

Next we use the symmetry property such that $\int_{-\infty}^{\infty}\omega'f_{s}(\omega')\mathrm{d}\omega'=0$ for any function $f_{s}(\omega')$ symmetric about $\omega'=0$ and the following expansions in $k/k'$ up to the first order
\begin{subequations}
\begin{equation}
P_{mp}(\mathbf{k}'+\frac{1}{2}\mathbf{k})=P_{mp}(\mathbf{k}')+\frac{k_{m}^{\prime}k_{p}^{\prime}k_{r}^{\prime}}{k^{\prime4}}k_{r}-\frac{k_{m}^{\prime}k_{p}+k_{m}k_{p}^{\prime}}{2k^{\prime2}}+\mathcal{O}\left(k^{2}\right),\label{eq:aux1-2}
\end{equation}
\begin{align}
P_{nqp}\left(\frac{1}{2}\mathbf{k}-\mathbf{k}'\right)=& -P_{nqp}\left(\mathbf{k}'\right)+2\frac{k_{n}^{\prime}k_{p}^{\prime}k_{q}^{\prime}k_{r}^{\prime}}{k^{\prime4}}k_{r}-\frac{k_{n}^{\prime}k_{q}^{\prime}k_{p}+k_{n}^{\prime}k_{p}^{\prime}k_{q}+2k_{n}k_{p}^{\prime}k_{q}^{\prime}}{2k^{\prime2}}\nonumber\\
&+\frac{1}{2}k_{q}P_{np}(\mathbf{k}')+\frac{1}{2}k_{p}P_{nq}(\mathbf{k}')+\mathcal{O}\left(k^{2}\right),\label{eq:aux2-2}
\end{align}
\begin{equation}
\frac{1}{\left|\gamma(\mathbf{k}'+\frac{1}{2}\mathbf{k},\omega')\right|^{2}}=\frac{1}{\omega^{\prime2}+k^{\prime4}+2k^{\prime2}k_{t}^{\prime}k_{t}}=\frac{1}{\omega^{\prime2}+k^{\prime4}}-\frac{2k^{\prime2}k_{t}^{\prime}k_{t}}{\left(\omega^{\prime2}+k^{\prime4}\right)^{2}}+\mathcal{O}\left(k^{2}\right),\label{eq:exp1}
\end{equation}
\begin{equation}
\Xi\left(\left|\mathbf{k}'+\frac{1}{2}\mathbf{k}\right|\right)=\Xi\left(k^{\prime}+\frac{1}{2k^{\prime}}\mathbf{k}^{\prime}\cdot\mathbf{k}+\mathcal{O}\left(k^{2}\right)\right)=\Xi\left(k^{\prime}\right)+\frac{1}{2k^{\prime}}k_{t}^{\prime}k_{t}\frac{\partial\Xi}{\partial k^{\prime}}+\mathcal{O}\left(k^{2}\right),\label{eq:exp2}
\end{equation}
\end{subequations}
which yields
\begin{align}
P_{imn}\int\mathrm{d}^{4}q'\left\langle \hat{u}_{m}^{>}(\mathbf{q}')\hat{u}_{n}^{>}(\mathbf{q}-\mathbf{q}')\right\rangle _{c}= & \,\qquad\qquad\qquad\qquad\qquad\qquad\qquad\qquad\qquad\qquad\qquad\qquad\qquad\qquad\qquad\nonumber \\
 & \hspace{-55mm}-2\mathrm{i}\epsilon P_{imn}\hat{u}_{p}^{<}(\mathbf{q})\int_{\lambda_{1}}^{\Lambda}\frac{\mathrm{d}k'}{k^{\prime2}}\int\mathrm{d}\mathring{\Omega}\int_{-\infty}^{\infty}\mathrm{d}\omega'\frac{\left[k^{\prime2}\Xi\left(k^{\prime}\right)-2k_{t}^{\prime}k_{t}\left(\Xi\left(k^{\prime}\right)-\frac{1}{4}k^{\prime}\frac{\partial\Xi}{\partial k^{\prime}}\right)\right]D_{npm}(\mathbf{k}',\mathbf{k})}{\left(\omega^{\prime2}+k^{\prime4}\right)^{2}}\nonumber \\
 &\hspace{-55mm} +\mathcal{O}\left(\epsilon^{2}\right),\label{eq:uu_corr-1-2-1-2-1}
\end{align}
with
\begin{equation}
D_{npm}(\mathbf{k}',\mathbf{k})=\left[-k_{p}^{\prime}P_{n3}\left(\mathbf{k}'\right)+\delta_{np}k_{z}-\frac{k_{p}^{\prime}k_{n}^{\prime}}{k^{\prime2}}k_{z}-\delta_{np}\frac{k_{z}^{\prime}k_{q}^{\prime}}{k^{\prime2}}k_{q}+\frac{k_{p}^{\prime}k_{z}^{\prime}}{k^{\prime2}}k_{n}\right]P_{m3}(\mathbf{k}')\label{D_npm}.
\end{equation}
Furthermore, by the use of
\begin{subequations} 
\begin{equation}
P_{imn}(\mathbf{k})\delta_{mn}=0,\qquad k_{p}\hat{u}_{p}^{<}(\mathbf{q})=0,\label{eq:Pimn_symmetries}
\end{equation}
\begin{equation}
\int\mathrm{d}\mathring{\varOmega}\underset{N}{\underbrace{k_{m}\dots k_{n}k_{k}}}=0,\quad\textrm{for any odd \emph{N} and all }m,\dots,n,\,k,\label{eq:odd_k_2}
\end{equation}
\begin{equation}
\int\mathrm{d}\mathring{\varOmega}\frac{k_{m}k_{n}}{k^{2}}=\frac{4\pi}{3}\delta_{mn}\label{eq:delta_formula_1-1}
\end{equation}
\begin{equation}
\int\frac{k_{j}k_{n}}{k^{2}}\cos^{2}\theta\mathrm{d}\mathring{\varOmega}=\frac{4\pi}{15}\left(\delta_{jn}+2\delta_{j3}\delta_{n3}\right),\label{eq:delta_formula_2-3}
\end{equation}
\begin{equation}
\int\mathrm{d}\mathring{\varOmega}\frac{k_{m}k_{n}k_{p}k_{q}}{k^{4}}=\frac{4\pi}{15}\left(\delta_{mn}\delta_{pq}+\delta_{mp}\delta_{nq}+\delta_{mq}\delta_{np}\right)\label{eq:delta_formula_2-1}
\end{equation}
\begin{align}
\int\frac{k_{t}k_{p}k_{m}k_{n}}{k^{4}}\cos^{2}\theta\mathrm{d}\mathring{\varOmega}=\,\, & \frac{4\pi}{105}\left(\delta_{tp}\delta_{mn}+\delta_{tm}\delta_{pn}+\delta_{tn}\delta_{pm}\right)\nonumber \\
 & +\frac{8\pi}{105}\left(\delta_{tp}\delta_{m3}\delta_{n3}+\delta_{tm}\delta_{p3}\delta_{n3}+\delta_{tn}\delta_{p3}\delta_{m3}+\delta_{pm}\delta_{t3}\delta_{n3}\right.\nonumber \\
 & \left.\qquad\quad+\delta_{pn}\delta_{t3}\delta_{m3}+\delta_{mn}\delta_{t3}\delta_{p3}\right)\label{eq:delta_formula_C_3}
\end{align}
\begin{equation}
\int_{-\infty}^{\infty}\frac{\mathrm{d}\omega'}{\left(\omega^{\prime2}+k^{\prime4}\right)^{2}}=\frac{\pi}{2k^{\prime6}},\label{eq:Int_3}
\end{equation}
\end{subequations}
where $\theta$ is the polar angle in spherical
coordinates $(k,\theta,\phi)$ one obtains
\begin{align}
-\frac{1}{2}\mathrm{i}\epsilon P_{imn}\int\mathrm{d}^{4}q'\left\langle \hat{u}_{m}^{>}(\mathbf{q}')\hat{u}_{n}^{>}(\mathbf{q}-\mathbf{q}')\right\rangle _{c} & \,\qquad\qquad\qquad\qquad\qquad\qquad\qquad\qquad\qquad\qquad\qquad\qquad\qquad\qquad\qquad\nonumber \\
 & \hspace{-62mm}=-\epsilon^{2}\frac{2\pi^{2}}{105}P_{imn}\hat{u}_{p}^{<}(\mathbf{q})\int_{\lambda_{1}}^{\Lambda}\frac{\mathrm{d}k'}{k^{\prime6}}\left\{ \left[4\Xi\left(k^{\prime}\right)-k^{\prime}\frac{\partial\Xi}{\partial k^{\prime}}\right]\delta_{pn}k_{m}+\left[8\Xi\left(k^{\prime}\right)+5k^{\prime}\frac{\partial\Xi}{\partial k^{\prime}}\right]\delta_{pn}\delta_{m3}k_{z}\right\} \nonumber \\
 & \hspace{-62mm}\;\;-\epsilon^{2}\frac{2\pi^{2}}{105}P_{imn}\hat{u}_{p}^{<}(\mathbf{q})\int_{\lambda_{1}}^{\Lambda}\frac{\mathrm{d}k'}{k^{\prime6}}\left[\Xi\left(k^{\prime}\right)+5k^{\prime}\frac{\partial\Xi}{\partial k^{\prime}}\right]\delta_{p3}\delta_{m3}k_{n}+\mathcal{O}\left(\epsilon^{3}\right)\nonumber \\
 & \hspace{-62mm}=-\epsilon^{2}\frac{2\pi^{2}}{105}\int_{\lambda_{1}}^{\Lambda}\frac{\mathrm{d}k'}{k^{\prime6}}\left[4\Xi\left(k^{\prime}\right)-k^{\prime}\frac{\partial\Xi}{\partial k^{\prime}}\right]k^{2}\hat{u}_{i}^{<}(\mathbf{q})-\epsilon^{2}\frac{2\pi^{2}}{105}\int_{\lambda_{1}}^{\Lambda}\frac{\mathrm{d}k'}{k^{\prime6}}\left[8\Xi\left(k^{\prime}\right)+5k^{\prime}\frac{\partial\Xi}{\partial k^{\prime}}\right]k_{z}^{2}\hat{u}_{i}^{<}(\mathbf{q})\nonumber \\
 & \hspace{-62mm}\;\;-\epsilon^{2}\frac{2\pi^{2}}{105}\int_{\lambda_{1}}^{\Lambda}\frac{\mathrm{d}k'}{k^{\prime6}}\left[\Xi\left(k^{\prime}\right)+5k^{\prime}\frac{\partial\Xi}{\partial k^{\prime}}\right]P_{i3}k^{2}\hat{u}_{z}^{<}(\mathbf{q})+\mathcal{O}\left(\epsilon^{3}\right)\label{eq:uu_corr-fin_1step}
\end{align}
We now utilize the assumption of narrowness of the first spectral
bite $\Lambda-\lambda_{1}=\delta\lambda\ll1$ and define the following
coefficients which describe the average effect of the short-wavelength
fluctuations with wave numbers from the narrow band $\lambda_{1}\leq k\leq\Lambda$
on the long-wavelength fluctuations corresponding to the band $0<k<\lambda_{1}$
(cf. (\ref{eq:uu_corr-fin_1step})) 
\begin{subequations}
\begin{equation}
\breve{\xi}\left(\lambda_{1}\right)=1+\epsilon^{2}\frac{2\pi^{2}}{105}\frac{\delta\lambda}{\lambda_{1}^{6}}\left[4\Xi\left(\lambda_{1}\right)-\lambda_{1}\frac{\partial\Xi}{\partial\lambda}\left(\lambda_{1}\right)\right],\label{eq:xi}
\end{equation}
\begin{equation}
\breve{\zeta}\left(\lambda_{1}\right)=\epsilon^{2}\frac{2\pi^{2}}{105}\frac{\delta\lambda}{\lambda_{1}^{6}}\left[8\Xi\left(\lambda_{1}\right)+5\lambda_{1}\frac{\partial\Xi}{\partial\lambda}\left(\lambda_{1}\right)\right],\label{eq:zeta}
\end{equation}
\begin{equation}
\breve{\chi}\left(\lambda_{1}\right)=\epsilon^{2}\frac{2\pi^{2}}{105}\frac{\delta\lambda}{\lambda_{1}^{6}}\left[\Xi\left(\lambda_{1}\right)+5\lambda_{1}\frac{\partial\Xi}{\partial\lambda}\left(\lambda_{1}\right)\right],\label{eq:chi}
\end{equation}
\end{subequations}
With the use of those definitions we can write
down the dynamical equation (\ref{eq:u<}) in the new form, with the
effect of the short-wavelength modes $\mathbf{u}^{>}$ expressed through
the effective Reynolds stresses (anisotropic turbulent viscosity)
\begin{align}
\left[-\mathrm{i}\omega+\breve{\xi}\left(\lambda_{1}\right)k^{2}+\breve{\zeta}\left(\lambda_{1}\right)k_{z}^{2}\right]\hat{u}_{i}^{<}(\mathbf{q})+\breve{\chi}\left(\lambda_{1}\right)P_{i3}k^{2}\hat{u}_{z}^{<}(\mathbf{q})= & \frac{P_{i3}}{k^{2}}\hat{Q}^{<}(\mathbf{q})-\frac{1}{2}\mathrm{i}\epsilon P_{imn}\mathbb{I}_{mn}^{(u^{<})}\left(\mathbf{q}\right).\label{eq:ren_eq_FS}
\end{align}
In order to proceed to the second step of the procedure we introduce
a short notation 
\begin{equation}
\breve{\gamma}=-\mathrm{i}\omega+\breve{\xi}\left(\lambda_{1}\right)k^{2}+\breve{\zeta}\left(\lambda_{1}\right)k_{z}^{2},\label{eq:gamma_new}
\end{equation}
which yields
\begin{equation}
\hat{u}_{i}^{<}(\mathbf{q})+\frac{\breve{\chi}\left(\lambda_{1}\right)k^{2}}{\breve{\gamma}}P_{i3}\hat{u}_{z}^{<}(\mathbf{q})=\frac{P_{i3}}{k^{2}\breve{\gamma}}\hat{Q}^{<}(\mathbf{q})-\frac{1}{2}\mathrm{i}\epsilon\frac{P_{imn}}{\breve{\gamma}}\mathbb{I}_{mn}^{(u^{<})}\left(\mathbf{q}\right)\overset{\mathrm{def}}{=}r.h.s._{i}.\label{eq:u<_after_first_step}
\end{equation}
Since
\begin{equation}
\hat{u}_{z}^{<}(\mathbf{q})=\frac{\mathbf{k}}{k_{z}}\cdot\left[\hat{\mathbf{e}}_{z}\times\left(\hat{\mathbf{e}}_{z}\times\hat{\mathbf{u}}^{<}(\mathbf{q})\right)\right],\label{eq:uz=00003D}
\end{equation}
we may take the cross-product of (\ref{eq:u<_after_first_step}) with
$\hat{\mathbf{e}}_{z}$ twice and then the dot-product with $\mathbf{k}$
to obtain
\begin{equation}
\hat{u}_{z}^{<}(\mathbf{q})=\frac{\breve{\gamma}}{\left(\breve{\gamma}+\breve{\chi}\left(\lambda_{1}\right)k_{h}^{2}\right)}\frac{\mathbf{k}}{k_{z}}\cdot\left[\hat{\mathbf{e}}_{z}\times\left(\hat{\mathbf{e}}_{z}\times\mathbf{r.h.s}\right)\right]=\frac{\breve{\gamma}}{\breve{\gamma}+\breve{\chi}\left(\lambda_{1}\right)k_{h}^{2}}r.h.s._{z},\label{eq:uz_expression}
\end{equation}
where $k_{h}^{2}=k^{2}-k_{z}^{2}$. This leads to
\begin{equation}
\hat{u}_{i}^{<}(\mathbf{q})=\frac{P_{i3}}{k^{2}\breve{\gamma}}\hat{Q}^{<}(\mathbf{q})-\frac{1}{2}\mathrm{i}\epsilon\frac{P_{imn}}{\breve{\gamma}}\mathbb{I}_{mn}^{(u^{<})}\left(\mathbf{q}\right)-\frac{\breve{\chi}\left(\lambda_{1}\right)k^{2}P_{i3}}{\breve{\gamma}+\breve{\chi}\left(\lambda_{1}\right)k_{h}^{2}}\left[\frac{1-\frac{k_{z}^{2}}{k^{2}}}{k^{2}\breve{\gamma}}\hat{Q}^{<}(\mathbf{q})-\frac{1}{2}\mathrm{i}\epsilon\frac{P_{3mn}}{\breve{\gamma}}\mathbb{I}_{mn}^{(u^{<})}\left(\mathbf{q}\right)\right],\label{eq:u<_after_first_step-1}
\end{equation}
and hence
\begin{equation}
\hat{u}_{i}^{<}(\mathbf{q})=\frac{P_{i3}}{k^{2}\breve{\Gamma}}\hat{Q}^{<}(\mathbf{q})-\frac{1}{2}\mathrm{i}\epsilon\frac{P_{imn}}{\breve{\gamma}}\mathbb{I}_{mn}^{(u^{<})}\left(\mathbf{q}\right)+\frac{1}{2}\mathrm{i}\epsilon\frac{\breve{\chi}\left(\lambda_{1}\right)k^{2}}{\breve{\Gamma}}\frac{P_{i3}P_{3mn}}{\breve{\gamma}}\mathbb{I}_{mn}^{(u^{<})}\left(\mathbf{q}\right),\label{eq:u<_after_first_step-2}
\end{equation}
where we have defined
\begin{align}
\breve{\Gamma}\overset{\mathrm{def}}{=}\breve{\gamma}+\breve{\chi}\left(\lambda_{1}\right)k_{h}^{2}= & -\mathrm{i}\omega+\left[\breve{\xi}\left(\lambda_{1}\right)+\breve{\chi}\left(\lambda_{1}\right)\right]k_{h}^{2}+\left[\breve{\xi}\left(\lambda_{1}\right)+\breve{\zeta}\left(\lambda_{1}\right)\right]k_{z}^{2}\nonumber \\
= & -\mathrm{i}\omega+\left[\breve{\xi}\left(\lambda_{1}\right)+\breve{\chi}\left(\lambda_{1}\right)\right]k^{2}+\left[\breve{\zeta}\left(\lambda_{1}\right)-\breve{\chi}\left(\lambda_{1}\right)\right]k_{z}^{2}.\label{eq:Gamma}
\end{align}

We now proceed to the next step of the iterative procedure which consists
of a step-by-step elimination of infinitesimally small wave-number
bands from the Fourier spectrum from the short-wavelength side. We
introduce $\lambda_{2}$, which satisfies
\begin{equation}
\delta\lambda=\lambda_{1}-\lambda_{2}\ll1,\label{eq:lambda_bite-1}
\end{equation}
and, again, split the remaining fluctuational Fourier spectrum $0\leq k\leq\lambda_{1}$
into two parts by defining new variables (but keeping the same notation)
\begin{equation}
\theta(k-\lambda_{2})\hat{u}_{i}^{<}(\mathbf{k},\omega)\mapsto\hat{u}_{i}^{>}(\mathbf{k},\omega),\qquad(\textrm{for }\lambda_{2}<k<\lambda_{1}),\label{eq:Fourier_division_1-1}
\end{equation}
\begin{equation}
\theta(\lambda_{2}-k)\hat{u}_{i}^{<}(\mathbf{k},\omega)\mapsto\hat{u}_{i}^{<}(\mathbf{k},\omega),\qquad(\textrm{for }k<\lambda_{2}),\label{eq:Fourier_division_2-1}
\end{equation}
and same way for $\hat{Q}$. The equations are also split, as in the
first step (cf. (\ref{eq:u<}) and (\ref{eq:u_gtr})), i.e. we have
\begin{align}
\left[-\mathrm{i}\omega+\breve{\xi}\left(\lambda_{1}\right)k^{2}+\breve{\zeta}\left(\lambda_{1}\right)k_{z}^{2}\right]\hat{u}_{i}^{<}(\mathbf{q})+\breve{\chi}\left(\lambda_{1}\right)P_{i3}k^{2}\hat{u}_{z}^{<}(\mathbf{q})= & \frac{P_{i3}}{k^{2}}\hat{Q}^{<}(\mathbf{q})-\frac{1}{2}\mathrm{i}\epsilon P_{imn}\mathbb{I}_{mn}^{(u^{<})}\left(\mathbf{q}\right)\qquad\qquad\qquad\nonumber \\
 & \hspace{-15mm}-\frac{1}{2}\mathrm{i}\epsilon P_{imn}\int d^{4}q'\left\langle \hat{u}_{m}^{>}(\mathbf{q}')\hat{u}_{n}^{>}(\mathbf{q}-\mathbf{q}')\right\rangle _{c},\label{eq:u<_second_step}
\end{align}
for the new long-wavelength modes and for the new short-wavelength
ones we get
\begin{align}
\hat{u}_{i}^{>}(\mathbf{q})= & \frac{P_{i3}}{k^{2}\breve{\Gamma}}\hat{Q}^{>}(\mathbf{q})-\mathrm{i}\epsilon\frac{P_{imn}}{\breve{\gamma}}\mathbb{J}_{mn}^{(u)}\left(\mathbf{q}\right)+\mathrm{i}\epsilon\frac{\breve{\chi}\left(\lambda_{1}\right)k^{2}}{\breve{\Gamma}}\frac{P_{i3}P_{3mn}}{\breve{\gamma}}\mathbb{J}_{mn}^{(u)}\left(\mathbf{q}\right)\nonumber \\
 & -\frac{1}{2}\mathrm{i}\epsilon\frac{P_{imn}}{\breve{\gamma}}\mathbb{I}_{mn}^{(u^{<})}\left(\mathbf{q}\right)+\frac{1}{2}\mathrm{i}\epsilon\frac{\breve{\chi}\left(\lambda_{1}\right)k^{2}}{\breve{\Gamma}}\frac{P_{i3}P_{3mn}}{\breve{\gamma}}\mathbb{I}_{mn}^{(u^{<})}\left(\mathbf{q}\right)+R_{i},\label{eq:u>_second_step}
\end{align}
where
\begin{equation}
\mathbb{J}_{mn}^{(u)}\left(\mathbf{q}\right)=\int d^{4}q''\hat{u}_{m}^{<}(\mathbf{q}'')\hat{u}_{n}^{>}(\mathbf{q}-\mathbf{q}''),\label{eq:Ju-2}
\end{equation}
\begin{equation}
R_{i}=-\frac{1}{2}\mathrm{i}\epsilon\left[\frac{P_{imn}}{\breve{\gamma}}-\frac{\breve{\chi}\left(\lambda_{1}\right)k^{2}}{\breve{\Gamma}}\frac{P_{i3}P_{3mn}}{\breve{\gamma}}\right]\mathbb{I}_{mn}^{(u^{>})}\left(\mathbf{q}\right),\label{eq:Ru-1}
\end{equation}
Of course now $\left\langle \cdot\right\rangle _{c}$ denotes conditional
average over the second shell ($\lambda_{2}\leq k\leq\lambda_{1}$)
statistical subensemble. 

Repetition of the sub-steps undertaken in the first step of the iterative
procedure, but with the modified expression for the short-wavelength
modes $\hat{u}_{i}^{>}(\mathbf{q})$, in general leads to a new expression
for the mean Reynolds stress. However, it will become clear, that
at the leading order the Reynolds stress remains uninfluenced by the
corrections $\breve{\xi}-1$, $\breve{\zeta}$ and $\breve{\chi}$,
which are all of the order $\epsilon^{2}$, thus unaltered with respect
to the previous step of the procedure (recall, that we neglect the
terms of the order $\mathcal{O}\left(\epsilon^{2}\right)$ in the fluctuational corrections $\left\langle \hat{u}_{m}^{>}\hat{u}_{n}^{>}\right\rangle _{c}$).
To demonstrate this explicitly we calculate
\begin{align}
\left\langle \hat{u}_{m}^{>}(\mathbf{q}')\hat{u}_{n}^{>}(\mathbf{q}-\mathbf{q}')\right\rangle _{c}= & \frac{P_{m3}\left(\mathbf{k}'\right)P_{n3}(\mathbf{k}-\mathbf{k}')}{k^{\prime2}\left|\mathbf{k}-\mathbf{k}'\right|^{2}\breve{\Gamma}\left(\mathbf{q}'\right)\breve{\Gamma}(\mathbf{q}-\mathbf{q}')}\left\langle \hat{Q}^{>}\left(\mathbf{q}'\right)\hat{Q}^{>}\left(\mathbf{q}-\mathbf{q}'\right)\right\rangle _{c}\nonumber \\
 &\hspace{-10mm} -\frac{\mathrm{i}\epsilon P_{m3}\left(\mathbf{k}'\right)}{k^{\prime2}\breve{\Gamma}(\mathbf{q}')}\frac{P_{npq}(\mathbf{k}-\mathbf{k}')}{\breve{\gamma}(\mathbf{q}-\mathbf{q}')}\left\langle \hat{Q}^{>}(\mathbf{q}')\mathbb{J}_{pq}^{(u)}\left(\mathbf{q}-\mathbf{q}'\right)\right\rangle _{c}\nonumber \\
 &\hspace{-10mm} -\frac{\mathrm{i}\epsilon P_{m3}\left(\mathbf{k}'\right)}{k^{\prime2}\breve{\Gamma}\left(\mathbf{q}'\right)}\frac{\breve{\chi}\left|\mathbf{k}-\mathbf{k}'\right|^{2}}{\breve{\Gamma}(\mathbf{q}-\mathbf{q}')}\frac{P_{n3}(\mathbf{k}-\mathbf{k}')P_{3pq}(\mathbf{k}-\mathbf{k}')}{\breve{\gamma}(\mathbf{q}-\mathbf{q}')}\left\langle \hat{Q}^{>}(\mathbf{q}')\mathbb{J}_{pq}^{(u)}\left(\mathbf{q}-\mathbf{q}'\right)\right\rangle _{c}\nonumber \\
 &\hspace{-10mm} -\frac{\mathrm{i}\epsilon P_{n3}(\mathbf{k}-\mathbf{k}')}{\left|\mathbf{k}-\mathbf{k}'\right|^{2}\breve{\Gamma}(\mathbf{q}-\mathbf{q}')}\frac{P_{mpq}\left(\mathbf{k}'\right)}{\breve{\gamma}\left(\mathbf{q}'\right)}\left\langle \hat{Q}^{>}\left(\mathbf{q}-\mathbf{q}'\right)\mathbb{J}_{pq}^{(u)}\left(\mathbf{q}'\right)\right\rangle _{c}\nonumber \\
 &\hspace{-10mm} -\frac{\mathrm{i}\epsilon P_{n3}(\mathbf{k}-\mathbf{k}')}{\left|\mathbf{k}-\mathbf{k}'\right|^{2}\breve{\Gamma}(\mathbf{q}-\mathbf{q}')}\frac{\breve{\chi}k^{\prime2}}{\breve{\Gamma}\left(\mathbf{q}'\right)}\frac{P_{m3}\left(\mathbf{k}'\right)P_{3pq}\left(\mathbf{k}'\right)}{\breve{\gamma}\left(\mathbf{q}'\right)}\left\langle \hat{Q}^{>}\left(\mathbf{q}-\mathbf{q}'\right)\mathbb{J}_{pq}^{(u)}\left(\mathbf{q}'\right)\right\rangle _{c}\nonumber \\
 &\hspace{-10mm} +\mathcal{O}\left(\epsilon^{2}\right).\label{eq:uu_corr-2}
\end{align}
The first term in (\ref{eq:uu_corr-2}) is proportional to $\left\langle \hat{Q}^{>}\left(\mathbf{q}'\right)\hat{Q}^{>}\left(\mathbf{q}-\mathbf{q}'\right)\right\rangle _{c}\sim\delta(\mathbf{k})\delta(\omega)$,
hence it does not contribute to the large-scale dynamics and it will
be disregarded, as in the first step. Substituting once again for
$\hat{\mathbf{u}}^{>}$ from (\ref{eq:u>_second_step}) into the $\mathbb{J}^{(u)}$-terms
in (\ref{eq:uu_corr-2}) and making use of the symmetry $\mathbf{q}'\mapsto\mathbf{q}-\mathbf{q}'$
under the integral $\int d^{4}q'$ one obtains
\begin{align}
\int\mathrm{d}^{4}q'\left\langle \hat{u}_{m}^{>}(\mathbf{q}')\hat{u}_{n}^{>}(\mathbf{q}-\mathbf{q}')\right\rangle _{c}= & \,\qquad\qquad\qquad\qquad\qquad\qquad\qquad\qquad\qquad\qquad\qquad\qquad\qquad\qquad\qquad\nonumber \\
 &\hspace{-53mm} -\mathrm{i}\epsilon\hspace{-1.5mm}\int\hspace{-1.5mm}\mathrm{d}^{4}q'\hspace{-1.5mm}\int\hspace{-1.5mm}\mathrm{d}^{4}q''\frac{\hat{u}_{p}^{<}(\mathbf{q}'')P_{m3}\left(\mathbf{k}'\right)P_{npq}(\mathbf{k}-\mathbf{k}')P_{q3}(\mathbf{k}-\mathbf{k}'-\mathbf{k}'')}{k^{\prime2}\left|\mathbf{k}-\mathbf{k}'-\mathbf{k}''\right|^{2}\breve{\Gamma}(\mathbf{q}')\breve{\Gamma}(\mathbf{q}-\mathbf{q}'-\mathbf{q}'')\breve{\gamma}(\mathbf{q}-\mathbf{q}')}\left\langle \hat{Q}^{>2}\right\rangle_{c}\nonumber \\
 &\hspace{-53mm} -\mathrm{i}\epsilon\breve{\chi}\hspace{-1.5mm}\int\hspace{-1.5mm}\mathrm{d}^{4}q'\hspace{-1.5mm}\int\hspace{-1.5mm}\mathrm{d}^{4}q''\frac{\hat{u}_{p}^{<}(\mathbf{q}'')\left|\mathbf{k}-\mathbf{k}'\right|^{2}P_{m3}\left(\mathbf{k}'\right)P_{n3}(\mathbf{k}-\mathbf{k}')P_{3pq}(\mathbf{k}-\mathbf{k}')P_{q3}(\mathbf{k}-\mathbf{k}'-\mathbf{k}'')}{k^{\prime2}\left|\mathbf{k}-\mathbf{k}'-\mathbf{k}''\right|^{2}\breve{\Gamma}\left(\mathbf{q}'\right)\breve{\Gamma}(\mathbf{q}-\mathbf{q}')\breve{\Gamma}(\mathbf{q}-\mathbf{q}'-\mathbf{q}'')\breve{\gamma}(\mathbf{q}-\mathbf{q}')}\left\langle \hat{Q}^{>2}\right\rangle_{c}\nonumber \\
 &\hspace{-53mm} +(m\leftrightarrow n)+\mathcal{O}\left(\epsilon^{2}\right),\label{eq:uu_corr-1-1}
\end{align}
where $(m\leftrightarrow n)$ in (\ref{eq:uu_corr-1-1}) denotes terms
of the same structure as the two previous ones but with exchanged
indices $m$ and $n$ and
\begin{equation}
\left\langle \hat{Q}^{>2}\right\rangle_{c}=\left\langle \hat{Q}^{>}(\mathbf{q}')\hat{Q}^{>}\left(\mathbf{q}-\mathbf{q}'-\mathbf{q}''\right)\right\rangle_{c}.\label{QQ}
\end{equation}
We can now substitute for the heat-source correlations
\begin{equation}
\left\langle \hat{Q}(\mathbf{k},\omega)\hat{Q}(\mathbf{k}',\omega')\right\rangle =\Xi\left(k\right)\delta(\mathbf{k}+\mathbf{k}')\delta(\omega+\omega'),\label{eq:force_correlations-1-1}
\end{equation}
cf. (\ref{eq:force_correlations}), into (\ref{eq:uu_corr-1-1}) and
perform the $\mathbf{q}''$ integral which yields
\begin{align}
-\frac{1}{2}\mathrm{i}\epsilon P_{imn}\int\mathrm{d}^{4}q'\left\langle \hat{u}_{m}^{>}(\mathbf{q}')\hat{u}_{n}^{>}(\mathbf{q}-\mathbf{q}')\right\rangle _{c}= & \,\qquad\qquad\qquad\qquad\qquad\qquad\qquad\qquad\qquad\qquad\qquad\qquad\qquad\qquad\qquad\nonumber \\
 &\hspace{-55mm} -\epsilon^{2}P_{imn}\hat{u}_{p}^{<}(\mathbf{q})\int\mathrm{d}^{4}q'\frac{\Xi\left(k'\right)P_{m3}\left(\mathbf{k}'\right)P_{npq}(\mathbf{k}-\mathbf{k}')P_{q3}(\mathbf{k}')}{k^{\prime4}\left|\breve{\Gamma}\left(\mathbf{q}'\right)\right|^{2}\breve{\gamma}(\mathbf{q}-\mathbf{q}')}\nonumber \\
 &\hspace{-55mm} -\epsilon^{2}\breve{\chi}P_{imn}\hat{u}_{p}^{<}(\mathbf{q})\int\mathrm{d}^{4}q'\frac{\Xi\left(k'\right)\left|\mathbf{k}-\mathbf{k}'\right|^{2}P_{m3}\left(\mathbf{k}'\right)P_{n3}(\mathbf{k}-\mathbf{k}')P_{3pq}(\mathbf{k}-\mathbf{k}')P_{q3}(\mathbf{k}')}{k^{\prime4}\left|\breve{\Gamma}\left(\mathbf{q}'\right)\right|^{2}\breve{\Gamma}(\mathbf{q}-\mathbf{q}')\breve{\gamma}(\mathbf{q}-\mathbf{q}')}\nonumber \\
 &\hspace{-55mm} +\mathcal{O}\left(\epsilon^{3}\right),\label{eq:uu_corr-1-1-1}
\end{align}
where\begin{subequations}
\begin{equation}
\breve{\gamma}=-\mathrm{i}\omega+\breve{\xi}k^{2}+\breve{\zeta}k_{z}^{2}=\gamma+\mathcal{O}\left(\epsilon^{2}\right),\label{eq:g1}
\end{equation}
\begin{equation}
\breve{\Gamma}=-\mathrm{i}\omega+\left(\breve{\xi}+\breve{\chi}\right)k^{2}+\left(\breve{\zeta}-\breve{\chi}\right)k_{z}^{2}=\gamma+\mathcal{O}\left(\epsilon^{2}\right),\label{eq:g2}
\end{equation}
\begin{equation}
\breve{\chi}=\mathcal{O}\left(\epsilon^{2}\right).\label{eq:g3}
\end{equation}
\end{subequations}
Hence we can write at leading order
\begin{align}
-\frac{1}{2}\mathrm{i}\epsilon P_{imn}\int\mathrm{d}^{4}q'\left\langle \hat{u}_{m}^{>}(\mathbf{q}')\hat{u}_{n}^{>}(\mathbf{q}-\mathbf{q}')\right\rangle _{c}= & \,\qquad\qquad\qquad\qquad\qquad\qquad\qquad\qquad\qquad\qquad\qquad\qquad\qquad\qquad\qquad\nonumber \\
 &\hspace{-15mm} -\epsilon^{2}P_{imn}\hat{u}_{p}^{<}(\mathbf{q})\int\mathrm{d}^{4}q'\frac{\Xi\left(k'\right)P_{m3}\left(\mathbf{k}'\right)P_{npq}(\mathbf{k}-\mathbf{k}')P_{q3}(\mathbf{k}')}{k^{\prime4}\left|\gamma\left(\mathbf{q}'\right)\right|^{2}\gamma(\mathbf{q}-\mathbf{q}')}\nonumber \\
 &\hspace{-15mm} +\mathcal{O}\left(\epsilon^{3}\right),\label{eq:uu_corr-1-1-1-2}
\end{align}
which is exactly the same as in the first step (cf. (\ref{eq:uu_corr-1-2}))
and therefore the resulting corrections must have the same form as
in (\ref{eq:ren_eq_FS}), (\ref{eq:xi}-c). It follows, that
\begin{subequations}
\begin{equation}
\breve{\xi}\left(\lambda_{2}\right)=\breve{\xi}\left(\lambda_{1}\right)+\epsilon^{2}\frac{2\pi^{2}}{105}\frac{\delta\lambda}{\lambda_{2}^{6}}\left[4\Xi\left(\lambda_{2}\right)-\lambda_{2}\frac{\partial\Xi}{\partial\lambda}\left(\lambda_{2}\right)\right],\label{eq:xi-2}
\end{equation}
\begin{equation}
\breve{\zeta}\left(\lambda_{2}\right)=\breve{\zeta}\left(\lambda_{1}\right)+\epsilon^{2}\frac{2\pi^{2}}{105}\frac{\delta\lambda}{\lambda_{2}^{6}}\left[8\Xi\left(\lambda_{2}\right)+5\lambda_{2}\frac{\partial\Xi}{\partial\lambda}\left(\lambda_{2}\right)\right],\label{eq:zeta-2}
\end{equation}
\begin{equation}
\breve{\chi}\left(\lambda_{2}\right)=\breve{\chi}\left(\lambda_{1}\right)+\epsilon^{2}\frac{2\pi^{2}}{105}\frac{\delta\lambda}{\lambda_{2}^{6}}\left[\Xi\left(\lambda_{2}\right)+5\lambda_{2}\frac{\partial\Xi}{\partial\lambda}\left(\lambda_{2}\right)\right].\label{eq:chi-2}
\end{equation}
\end{subequations}
If we now return to the dimensional units (recall,
that we had chosen $L^{2}/\nu$ for the time scale, $L$ for the spatial
scale and $\kappa\nu U/g\bar{\alpha}L^{4}$ for the heat source scale;
the latter implies $\kappa^{2}\nu U^{2}/g^{2}\bar{\alpha}^{2}L^{3}$
for the scale of $\Xi(k)$)
\begin{align}
\left[-\mathrm{i}\omega+\nu\breve{\xi}\left(\lambda_{1}\right)k^{2}+\nu\breve{\zeta}\left(\lambda_{1}\right)k_{z}^{2}\right]\hat{u}_{i}^{<}(\mathbf{q})+\nu\breve{\chi}\left(\lambda_{1}\right)P_{i3}k^{2}\hat{u}_{z}^{<}(\mathbf{q})= & \frac{g\bar{\alpha}}{\kappa}\frac{P_{i3}}{k^{2}}\hat{Q}^{<}(\mathbf{q})-\frac{1}{2}\mathrm{i}P_{imn}\mathbb{I}_{mn}^{(u^{<})}\left(\mathbf{q}\right),\label{eq:ren_eq_FS-2}
\end{align}
we obtain
\begin{subequations}
\begin{equation}
\frac{\breve{\xi}\left(\lambda_{1}\right)-\breve{\xi}\left(\lambda_{2}\right)}{\delta\lambda}=-\frac{2\pi^{2}}{105}\frac{g^{2}\bar{\alpha}^{2}}{\nu^{3}\kappa^{2}}\frac{1}{\lambda_{2}^{6}}\left[4\Xi\left(\lambda_{2}\right)-\lambda_{2}\frac{\partial\Xi}{\partial\lambda}\left(\lambda_{2}\right)\right],\label{eq:xi-2-1}
\end{equation}
\begin{equation}
\frac{\breve{\zeta}\left(\lambda_{1}\right)-\breve{\zeta}\left(\lambda_{2}\right)}{\delta\lambda}=-\frac{2\pi^{2}}{105}\frac{g^{2}\bar{\alpha}^{2}}{\nu^{3}\kappa^{2}}\frac{1}{\lambda_{2}^{6}}\left[8\Xi\left(\lambda_{2}\right)+5\lambda_{2}\frac{\partial\Xi}{\partial\lambda}\left(\lambda_{2}\right)\right],\label{eq:zeta-2-1}
\end{equation}
\begin{equation}
\frac{\breve{\chi}\left(\lambda_{1}\right)-\breve{\chi}\left(\lambda_{2}\right)}{\delta\lambda}=-\frac{2\pi^{2}}{105}\frac{g^{2}\bar{\alpha}^{2}}{\nu^{3}\kappa^{2}}\frac{1}{\lambda_{2}^{6}}\left[\Xi\left(\lambda_{2}\right)+5\lambda_{2}\frac{\partial\Xi}{\partial\lambda}\left(\lambda_{2}\right)\right].\label{eq:chi-2-1}
\end{equation}
\end{subequations}
It is now clear, that in all the following steps
of the iterative, asymptotic procedure no terms with new structure
can appear in the velocity equation and thus we can now take the continuous
limit $\delta\lambda\rightarrow0$ of the obtained recursions. Let
us introduce the following simple form of the heat source correlation
function (note, that in an isotropic case driven by a random forcing
$\mathbf{f}$, a scaling of the form $\left\langle f_{i}f_{j}\right\rangle \sim k^{2}$
was shown by Lifshitz and Pitaevskii (1987) \cite{LifshitzPitaevskii1987} to describe systems in thermal
equilibrium thus to study non-equilibrium flows we consider forcing
with significantly smaller scaling exponent)
\begin{equation}
\Xi\left(k\right)=\frac{\mathbb{Q}^{2}L^{3}}{\nu k^{2}},\label{eq:Xi}
\end{equation}
which ensures, that the spectral density of the heat source $Q^{2}$
\begin{equation}
\int_{0}^{\Lambda}k^{2}\mathrm{d}k\int\mathrm{d}\mathring{\varOmega}_{\mathbf{k}}\int\mathrm{d}^{4}q^{\prime}\left\langle \hat{Q}(\mathbf{k},\omega)\hat{Q}(\mathbf{k}',\omega')\right\rangle =\int_{0}^{\Lambda}\frac{4\pi\mathbb{Q}^{2}L^{3}}{\nu}\mathrm{d}k,\label{eq:F2-1}
\end{equation}
is uniform and where $\mathbb{Q}$ is the magnitude of the heat delivery
rate (in $K/s$); this yields
\begin{subequations}
\begin{equation}
\frac{\mathrm{d}\breve{\xi}}{\mathrm{d}\lambda}=-\frac{4\pi^{2}}{35}\frac{g^{2}\bar{\alpha}^{2}L^{3}}{\nu^{4}\kappa^{2}}\frac{\mathbb{Q}^{2}}{\lambda^{8}}\quad\Rightarrow\quad\breve{\xi}(\Lambda)-\breve{\xi}(\lambda)=-\frac{4\pi^{2}}{245}\frac{g^{2}\bar{\alpha}^{2}L^{3}\mathbb{Q}^{2}}{\nu^{4}\kappa^{2}}\left(\frac{1}{\lambda^{7}}-\frac{1}{\Lambda^{7}}\right),\label{eq:xi-2-1-1}
\end{equation}
\begin{equation}
\frac{\mathrm{d}\breve{\zeta}}{\mathrm{d}\lambda}=\frac{4\pi^{2}}{105}\frac{g^{2}\bar{\alpha}^{2}L^{3}}{\nu^{4}\kappa^{2}}\frac{\mathbb{Q}^{2}}{\lambda^{8}}\quad\Rightarrow\quad\breve{\zeta}(\Lambda)-\breve{\zeta}(\lambda)=\frac{4\pi^{2}}{735}\frac{g^{2}\bar{\alpha}^{2}L^{3}\mathbb{Q}^{2}}{\nu^{4}\kappa^{2}}\left(\frac{1}{\lambda^{7}}-\frac{1}{\Lambda^{7}}\right),\label{eq:zeta-2-1-1}
\end{equation}
\begin{equation}
\frac{\mathrm{d}\breve{\chi}}{\mathrm{d}\lambda}=\frac{6\pi^{2}}{35}\frac{g^{2}\bar{\alpha}^{2}L^{3}}{\nu^{4}\kappa^{2}}\frac{\mathbb{Q}^{2}}{\lambda^{8}}\quad\Rightarrow\quad\breve{\chi}(\Lambda)-\breve{\chi}(\lambda)=\frac{6\pi^{2}}{245}\frac{g^{2}\bar{\alpha}^{2}L^{3}\mathbb{Q}^{2}}{\nu^{4}\kappa^{2}}\left(\frac{1}{\lambda^{7}}-\frac{1}{\Lambda^{7}}\right).\label{eq:chi-2-1-1}
\end{equation}
\end{subequations}
Application of the ``initial'' conditions $\breve{\xi}(\Lambda)=1$,
$\breve{\zeta}(\Lambda)=0$, $\breve{\chi}(\Lambda)=0$ and the limit
$\lambda=k_{\ell}=2\pi/\ell\ll\Lambda$ leads to
\begin{subequations}
\begin{equation}
\breve{\xi}(k_{\ell})=1+\frac{4\pi^{2}}{245}\frac{g^{2}\bar{\alpha}^{2}L^{3}\mathbb{Q}^{2}}{\nu^{4}\kappa^{2}k_{\ell}^{7}},\label{eq:xi-1-1-1}
\end{equation}
\begin{equation}
\breve{\zeta}(k_{\ell})=-\frac{4\pi^{2}}{735}\frac{g^{2}\bar{\alpha}^{2}L^{3}\mathbb{Q}^{2}}{\nu^{4}\kappa^{2}k_{\ell}^{7}},\label{eq:zeta-1-1-1}
\end{equation}
\begin{equation}
\breve{\chi}(k_{\ell})=-\frac{6\pi^{2}}{245}\frac{g^{2}\bar{\alpha}^{2}L^{3}\mathbb{Q}^{2}}{\nu^{4}\kappa^{2}k_{\ell}^{7}}.\label{eq:chi-1-1-1}
\end{equation}
\end{subequations}
where $\ell$ can be thought of as the length-scale
of most energetic eddies in the turbulent flow. On defining
\begin{subequations}
\begin{equation}
\varsigma=\frac{2\pi^{2}}{735}\frac{g^{2}\bar{\alpha}^{2}L^{3}\mathbb{Q}^{2}}{\nu^{3}\kappa^{2}k_{\ell}^{7}},\label{eq:varsigma}
\end{equation}
\begin{equation}
\xi=\nu\breve{\xi}(k_{\ell})=\nu+6\varsigma,\label{eq:xi-1-1-1-1}
\end{equation}
\begin{equation}
\zeta=-\nu\breve{\zeta}(k_{\ell})=2\varsigma,\label{eq:zeta-1-1-1-1}
\end{equation}
\begin{equation}
\chi=-\nu\breve{\chi}(k_{\ell})=9\varsigma.\label{eq:chi-1-1-1-1}
\end{equation}
\end{subequations}
the large-scale flow is governed by
\begin{equation}
\frac{\partial\left\langle \mathbf{U}\right\rangle }{\partial t}+\left(\left\langle \mathbf{U}\right\rangle \cdot\nabla\right)\left\langle \mathbf{U}\right\rangle =-\nabla\left\langle \Pi\right\rangle -9\varsigma\nabla^{2}\left\langle U\right\rangle _{z}\hat{\mathbf{e}}_{z}+\left(\nu+6\varsigma\right)\nabla^{2}\left\langle \mathbf{U}\right\rangle -2\varsigma\frac{\partial^{2}\left\langle \mathbf{U}\right\rangle }{\partial z^{2}},\label{eq:Ren_NS-1-1}
\end{equation}
or
\begin{equation}
\frac{\partial\left\langle \mathbf{U}\right\rangle }{\partial t}+\left(\left\langle \mathbf{U}\right\rangle \cdot\nabla\right)\left\langle \mathbf{U}\right\rangle =-\nabla\left\langle \Pi\right\rangle -9\varsigma\nabla^{2}\left\langle U\right\rangle _{z}\hat{\mathbf{e}}_{z}+\left(\nu+6\varsigma\right)\nabla_{h}^{2}\left\langle \mathbf{U}\right\rangle +\left(\nu+4\varsigma\right)\frac{\partial^{2}\left\langle \mathbf{U}\right\rangle }{\partial z^{2}}.\label{eq:MF_eq_WN}
\end{equation}

\section{Comments on full renormalization}

Full renormalization of the anisotropic problem at hand is not possible,
because the integrals in (\ref{eq:uu_corr-1-1-1}) without the simplification
(\ref{eq:g1}-c) coming from neglection of terms higher order in $\epsilon$
cannot be evaluated analytically. Moreover, as it will be demonstrated
below, the full integrals (\ref{eq:uu_corr-1-1-1}) (not approximated
to leading order in $\epsilon$) lead to introduction of yet another
term in the equation of the form $\sim P_{i3}k_{z}^{2}\hat{u}_{z}^{<}$,
thus making the further steps even more complicated; it can be demonstrated
however, that further steps of the renormalization procedure do not
introduce terms of new structure. Therefore the most general form
of the mean flow equation is the following
\begin{equation}
\frac{\partial\left\langle \mathbf{U}\right\rangle }{\partial t}+\left(\left\langle \mathbf{U}\right\rangle \cdot\nabla\right)\left\langle \mathbf{U}\right\rangle =-\nabla\left\langle \Pi\right\rangle -\chi\nabla^{2}\left\langle U\right\rangle _{z}\hat{\mathbf{e}}_{z}+\varrho\frac{\partial^{2}\left\langle U\right\rangle _{z}}{\partial z^{2}}\hat{\mathbf{e}}_{z}+\xi\nabla_{h}^{2}\left\langle \mathbf{U}\right\rangle +\zeta\frac{\partial^{2}\left\langle \mathbf{U}\right\rangle }{\partial z^{2}},\label{eq:gen_form_MF}
\end{equation}
where based on our weakly nonlinear results we may suppose, that $\xi>0$,
$\zeta>0$ and $\chi>0$, but the sign of the coefficient $\varrho$
remains undetermined. Within the weakly nonlinear approach this coefficient
is negligibly small $\varrho=\mathcal{O}\left(\epsilon^{4}\right)$,
however in fully developed strong turbulence it might be of comparable
magnitude with all the remaining coefficients. 

We will now demonstrate, that indeed, the renormalization leads to
the above general form of the mean flow equation. To that end we need
to return to the second step of the procedure and consider the full
integrals in (\ref{eq:uu_corr-1-1-1}). Making the aforementioned
substitution $\mathbf{k}'\rightarrow\mathbf{k}'+\frac{1}{2}\mathbf{k}$
to symmetrize the domain of integration one obtains
\begin{align}
-\frac{1}{2}\mathrm{i}\epsilon P_{imn}\int\mathrm{d}^{4}q'\left\langle \hat{u}_{m}^{>}(\mathbf{q}')\hat{u}_{n}^{>}(\mathbf{q}-\mathbf{q}')\right\rangle _{c}= & \,\qquad\qquad\qquad\qquad\qquad\qquad\qquad\qquad\qquad\qquad\qquad\qquad\qquad\qquad\qquad\nonumber \\
 &\hspace{-68mm} -\epsilon^{2}\frac{\mathbb{Q}^{2}L^{3}}{\nu}P_{imn}\hat{u}_{p}^{<}(\mathbf{q})\hspace{-1.5mm}\int\hspace{-1.5mm}\mathrm{d}k'\hspace{-1.5mm}\int\hspace{-1.5mm}\mathrm{d}\mathring{\Omega}\hspace{-1.5mm}\int_{-\infty}^{\infty}\hspace{-2mm}\mathrm{d}\omega'\frac{\mathbb{P}_{mq}\left(\mathbf{k}'+\frac{1}{2}\mathbf{k}\right)P_{npq}(\frac{1}{2}\mathbf{k}-\mathbf{k}')}{\left(k^{\prime2}+k_{r}^{\prime}k_{r}\right)^{2}\left|\breve{\Gamma}(\mathbf{k}'+\frac{1}{2}\mathbf{k},\omega')\right|^{2}\breve{\gamma}(\mathbf{k}'-\frac{1}{2}\mathbf{k},-\omega')}\nonumber \\
 &\hspace{-68mm} -\epsilon^{2}\frac{\mathbb{Q}^{2}L^{3}}{\nu}\breve{\chi}P_{imn}\hat{u}_{p}^{<}(\mathbf{q})\hspace{-1.5mm}\int\hspace{-1.5mm}\mathrm{d}k'\hspace{-1.5mm}\int\hspace{-1.5mm}\mathrm{d}\mathring{\Omega}\hspace{-1.5mm}\int_{-\infty}^{\infty}\hspace{-2mm}\mathrm{d}\omega'\frac{\left|\mathbf{k}'-\frac{1}{2}\mathbf{k}\right|^{2}\mathbb{P}_{mq}\left(\mathbf{k}'+\frac{1}{2}\mathbf{k}\right)P_{n3}(\mathbf{k}'-\frac{1}{2}\mathbf{k})P_{3pq}(\frac{1}{2}\mathbf{k}-\mathbf{k}')}{k^{\prime4}\left|\breve{\Gamma}\left(\mathbf{k}'+\frac{1}{2}\mathbf{k},\omega'\right)\right|^{2}\breve{\Gamma}(\mathbf{k}'-\frac{1}{2}\mathbf{k},-\omega')\breve{\gamma}(\mathbf{k}'-\frac{1}{2}\mathbf{k},-\omega')}\nonumber \\
 &\hspace{-68mm} +\mathcal{O}\left(\epsilon^{3}\right),\label{eq:uu_corr-1-1-1-1}
\end{align}
where
\begin{equation}
\mathbb{P}_{mq}\left(\mathbf{k}'+\frac{1}{2}\mathbf{k}\right)=P_{m3}\left(\mathbf{k}'+\frac{1}{2}\mathbf{k}\right)P_{q3}\left(\mathbf{k}'+\frac{1}{2}\mathbf{k}\right).\label{PPmq}
\end{equation}
Expansion in $k/k'$ up to the first order leads to
\begin{align}
-\frac{1}{2}\mathrm{i}\epsilon P_{imn}\int\mathrm{d}^{4}q'\left\langle \hat{u}_{m}^{>}(\mathbf{q}')\hat{u}_{n}^{>}(\mathbf{q}-\mathbf{q}')\right\rangle _{c}= & \,\qquad\qquad\qquad\qquad\qquad\qquad\qquad\qquad\qquad\qquad\qquad\qquad\qquad\qquad\qquad\nonumber \\
 &\hspace{-68mm} -\epsilon^{2}\frac{\mathbb{Q}^{2}L^{3}}{\nu}P_{imn}\hat{u}_{p}^{<}(\mathbf{q})\hspace{-1.5mm}\int\frac{\mathrm{d}k'}{k^{\prime}{}^{8}}\hspace{-1.5mm}\int\hspace{-1.5mm}\mathrm{d}\mathring{\Omega}\hspace{-1.5mm}\int_{-\infty}^{\infty}\hspace{-2mm}\mathrm{d}\omega'\frac{\mathcal{F}\left(\mathbf{k}',\omega'\right)\mathbb{P}_{mq}\left(\mathbf{k}'+\frac{1}{2}\mathbf{k}\right)P_{npq}(\frac{1}{2}\mathbf{k}-\mathbf{k}')}{\left|\breve{\Gamma}(\mathbf{k}'+\frac{1}{2}\mathbf{k},\omega')\right|^{2}\left|\breve{\gamma}(\mathbf{k}'-\frac{1}{2}\mathbf{k},\omega')\right|^{2}}\nonumber \\
 &\hspace{-68mm} -\epsilon^{2}\frac{\mathbb{Q}^{2}L^{3}}{\nu}\breve{\chi}P_{imn}\hat{u}_{p}^{<}(\mathbf{q})\hspace{-1.5mm}\int\frac{\mathrm{d}k'}{k^{\prime8}}\hspace{-1.5mm}\int\hspace{-1.5mm}\mathrm{d}\mathring{\Omega}\hspace{-1.5mm}\int_{-\infty}^{\infty}\hspace{-2mm}\mathrm{d}\omega'\frac{\mathcal{G}\left(\mathbf{k}',\omega'\right)\mathbb{P}_{mq}\left(\mathbf{k}'+\frac{1}{2}\mathbf{k}\right)P_{n3}(\mathbf{k}'-\frac{1}{2}\mathbf{k})P_{3pq}(\frac{1}{2}\mathbf{k}-\mathbf{k}')}{\left|\breve{\Gamma}\left(\mathbf{k}',\omega'\right)\right|^{4}\left|\breve{\gamma}(\mathbf{k}'-\frac{1}{2}\mathbf{k},\omega')\right|^{2}}\nonumber \\
 &\hspace{-68mm} +\mathcal{O}\left(\epsilon^{3}\right),\label{eq:uu_corr-1-1-1-1-1}
\end{align}
with
\begin{align}
\mathcal{F}\left(\mathbf{k}',\omega'\right)
\approx & \left[\breve{\xi}k^{\prime2}-\breve{\zeta}k_{z}^{\prime2}-\breve{\xi}k_{r}^{\prime}k_{r}+\breve{\zeta}k_{z}^{\prime}k_{z}\right]\left(k^{\prime2}-k_{r}^{\prime}k_{r}\right)^{2}\nonumber \\
\approx & k^{\prime4}\left[k^{\prime2}\left(\breve{\xi}-\breve{\zeta}X^{2}\right)+\breve{\zeta}k_{z}^{\prime}k_{z}+\left(2\breve{\zeta}X^{2}-3\breve{\xi}\right)k_{r}^{\prime}k_{r}\right]\label{eq:Fcal}
\end{align}
\begin{align}
\mathcal{G}\left(\mathbf{k}',\omega'\right)
\approx & \left\{ -\omega^{\prime2}+k^{\prime4}\left(\breve{\xi}-\breve{\zeta}X^{2}\right)\left[\left(\breve{\xi}-\breve{\chi}\right)-\left(\breve{\zeta}-\breve{\chi}\right)X^{2}\right]\right.\nonumber \\
 & \quad-\left[2\breve{\xi}\left(\breve{\xi}-\breve{\chi}\right)-\left(2\breve{\xi}\breve{\zeta}-\breve{\chi}\left(\breve{\xi}+\breve{\zeta}\right)\right)X^{2}\right]k^{\prime2}k_{r}^{\prime}k_{r}\nonumber \\
 & \left.\quad+\left[2\breve{\xi}\breve{\zeta}-\breve{\chi}\left(\breve{\xi}+\breve{\zeta}\right)-2\breve{\zeta}\left(\breve{\zeta}-\breve{\chi}\right)X^{2}\right]k^{\prime2}k_{z}^{\prime}k_{z}\right\} \left(k^{\prime2}-k_{r}^{\prime}k_{r}\right)^{3}\nonumber \\
\approx & -\omega^{\prime2}k^{\prime6}+k^{\prime10}\left(\breve{\xi}-\breve{\zeta}X^{2}\right)\left[\left(\breve{\xi}-\breve{\chi}\right)-\left(\breve{\zeta}-\breve{\chi}\right)X^{2}\right]\nonumber \\
 & +\left\{ 3\omega^{\prime2}-\left[5\breve{\xi}\left(\breve{\xi}-\breve{\chi}\right)-4\left(2\breve{\xi}\breve{\zeta}-\breve{\chi}\left(\breve{\xi}+\breve{\zeta}\right)\right)X^{2}+3\breve{\zeta}\left(\breve{\zeta}-\breve{\chi}\right)X^{4}\right]k^{\prime4}\right\} k^{\prime4}k_{r}^{\prime}k_{r}\nonumber \\
 & +\left[2\breve{\xi}\breve{\zeta}-\breve{\chi}\left(\breve{\xi}+\breve{\zeta}\right)-2\breve{\zeta}\left(\breve{\zeta}-\breve{\chi}\right)X^{2}\right]k^{\prime8}k_{z}^{\prime}k_{z}\label{eq:Gcal}
\end{align}
\begin{align}
 & \frac{1}{\left|\breve{\gamma}(\mathbf{k}'-\frac{1}{2}\mathbf{k},\omega')\right|^{2}}\frac{1}{\left|\breve{\Gamma}(\mathbf{k}'+\frac{1}{2}\mathbf{k},\omega')\right|^{2}}\approx\nonumber \\
 & \frac{1}{\left[\omega^{\prime2}+k^{\prime4}\left(\breve{\xi}-\breve{\zeta}X^{2}\right)^{2}\right]\left[\omega^{\prime2}+k^{\prime4}\left(\left(\breve{\xi}-\breve{\chi}\right)-\left(\breve{\zeta}-\breve{\chi}\right)X^{2}\right)^{2}\right]}\nonumber \\
 & +2\breve{\chi}k^{\prime2}\frac{\omega^{\prime2}\left\{ \left[\breve{\xi}\left(2-X^{2}\right)-\breve{\zeta}X^{2}-\breve{\chi}\left(1-X^{2}\right)\right]k_{r}^{\prime}k_{r}+\left[\breve{\zeta}\left(2X^{2}-1\right)-\breve{\xi}+\breve{\chi}\left(1-X^{2}\right)\right]k_{z}^{\prime}k_{z}\right\}}{\left[\omega^{\prime2}+k^{\prime4}\left(\left(\breve{\xi}-\breve{\chi}\right)-\left(\breve{\zeta}-\breve{\chi}\right)X^{2}\right)^{2}\right]^{2}\left[\omega^{\prime2}+k^{\prime4}\left(\breve{\xi}-\breve{\zeta}X^{2}\right)^{2}\right]^{2}}\nonumber\\
 & +2\breve{\chi}k^{\prime2}\frac{ k^{\prime4}\left(\breve{\xi}-\breve{\zeta}\right)\left(\breve{\xi}-\breve{\zeta}X^{2}\right)\left(\left(\breve{\xi}-\breve{\chi}\right)-\left(\breve{\zeta}-\breve{\chi}\right)X^{2}\right)\left(X^{2}k_{r}^{\prime}k_{r}-k_{z}^{\prime}k_{z}\right)}{\left[\omega^{\prime2}+k^{\prime4}\left(\left(\breve{\xi}-\breve{\chi}\right)-\left(\breve{\zeta}-\breve{\chi}\right)X^{2}\right)^{2}\right]^{2}\left[\omega^{\prime2}+k^{\prime4}\left(\breve{\xi}-\breve{\zeta}X^{2}\right)^{2}\right]^{2}},\label{eq:gamma_exp}
\end{align}
and under integration over the azimuthal angle $\int_{0}^{2\pi}\mathrm{d}\varphi$
different terms in the integrands transform into
\begin{align}
 & P_{imn}\hat{u}_{p}^{<}(\mathbf{q})f(X^{2})P_{m3}\left(\mathbf{k}'+\frac{1}{2}\mathbf{k}\right)P_{npq}(\frac{1}{2}\mathbf{k}-\mathbf{k}')P_{q3}(\mathbf{k}'+\frac{1}{2}\mathbf{k})\nonumber \\
 & \rightarrow P_{imn}\hat{u}_{p}^{<}(\mathbf{q})f(X^{2})\left[-\delta_{np}\frac{k_{z}^{\prime}k_{q}^{\prime}}{k^{\prime2}}k_{q}+\delta_{np}k_{z}-\frac{k_{p}^{\prime}k_{n}^{\prime}}{k^{\prime2}}k_{z}+\frac{k_{p}^{\prime}k_{z}^{\prime}}{k^{\prime2}}k_{n}\right]P_{m3}(\mathbf{k}')\nonumber \\
 & \rightarrow\pi f(X^{2})\left(1-X^{2}\right)^{2}k_{z}^{2}\hat{u}_{i}^{<}(\mathbf{q})+3\pi f(X^{2})X^{2}\left(1-X^{2}\right)k^{2}P_{i3}\hat{u}_{z}^{<}(\mathbf{q})\nonumber\\
 &\quad\;+\pi f(X^{2})\left[\frac{35}{2}X^{4}-21X^{2}+2\right]k_{z}^{2}P_{i3}\hat{u}_{z}^{<}(\mathbf{q}),\label{eq:aux1}
\end{align}
\begin{align}
 & P_{imn}\hat{u}_{p}^{<}(\mathbf{q})f(X^{2})P_{m3}\left(\mathbf{k}'+\frac{1}{2}\mathbf{k}\right)P_{n3}(\mathbf{k}'-\frac{1}{2}\mathbf{k})P_{3pq}(\frac{1}{2}\mathbf{k}-\mathbf{k}')P_{q3}(\mathbf{k}'+\frac{1}{2}\mathbf{k})\nonumber \\
 & \rightarrow P_{imn}\hat{u}_{p}^{<}(\mathbf{q})f(X^{2})\left[P_{m3}(\mathbf{k}')P_{n3}(\mathbf{k}')P_{p3}\left(\mathbf{k}'\right)k_{z}-P_{m3}(\mathbf{k}')P_{n3}(\mathbf{k}')P_{p3}\left(\mathbf{k}'\right)\frac{k_{z}^{\prime}k_{q}^{\prime}}{k^{\prime2}}k_{q}\right]\nonumber \\
 & \rightarrow\frac{\pi}{2}f(X^{2})X^{4}\left(1-X^{2}\right)^{2}k^{2}\hat{u}_{i}^{<}(\mathbf{q})\nonumber \\
 & \quad+\frac{\pi}{2}f(X^{2})X^{2}\left(4-17X^{2}+18X^{4}-5X^{6}\right)k_{z}^{2}\hat{u}_{i}^{<}(\mathbf{q})\nonumber \\
 & \quad+\frac{\pi}{2}f(X^{2})X^{2}\left(4-9X^{2}+10X^{4}-5X^{6}\right)P_{i3}k^{2}\hat{u}_{z}^{<}(\mathbf{q})\nonumber \\
 & \quad+\frac{\pi}{2}f(X^{2})\left(8-56X^{2}+123X^{4}-110X^{6}+35X^{8}\right)P_{i3}k_{z}^{2}\hat{u}_{z}^{<}(\mathbf{q}),\label{eq:aux2}
\end{align}
\begin{align}
& P_{imn}\hat{u}_{p}^{<}(\mathbf{q})f(X^{2})k_{p}^{\prime}P_{n3}\left(\mathbf{k}'\right)P_{m3}(\mathbf{k}')k_{z}^{\prime}k_{z}\nonumber\\
&\rightarrow  P_{imn}k_{z}\hat{u}_{p}^{<}(\mathbf{q})k^{\prime2}f(X^{2})\left(\delta_{m3}\delta_{n3}\frac{k_{z}^{\prime}k_{p}^{\prime}}{k^{\prime2}}-2X^{2}\delta_{m3}\frac{k_{n}^{\prime}k_{p}^{\prime}}{k^{\prime2}}+X^{2}\frac{k_{m}^{\prime}k_{n}^{\prime}k_{p}^{\prime}k_{z}^{\prime}}{k^{\prime4}}\right)\nonumber \\
&\rightarrow  2\pi k^{\prime2}f(X^{2})X^{2}\left(5X^{4}-9X^{2}+4\right)P_{i3}k_{z}^{2}\hat{u}_{z}^{<}(\mathbf{q})-2\pi k^{\prime2}f(X^{2})X^{2}\left(1-X^{2}\right)^{2}k_{z}^{2}\hat{u}_{i}^{<}(\mathbf{q}),\label{eq:aux3}
\end{align}
\begin{align}
& P_{imn}\hat{u}_{p}^{<}(\mathbf{q})f(X^{2})k_{p}^{\prime}P_{n3}\left(\mathbf{k}'\right)P_{m3}(\mathbf{k}')k_{r}^{\prime}k_{r}\nonumber\\
& \rightarrow  k_{r}P_{imn}\hat{u}_{p}^{<}(\mathbf{q})k^{\prime2}f(X^{2})\left(\delta_{n3}\delta_{m3}\frac{k_{p}^{\prime}k_{r}^{\prime}}{k^{\prime2}}-2\delta_{m3}\frac{k_{p}^{\prime}k_{r}^{\prime}k_{n}^{\prime}k_{z}^{\prime}}{k^{\prime4}}+X^{2}\frac{k_{p}^{\prime}k_{r}^{\prime}k_{m}^{\prime}k_{n}^{\prime}}{k^{\prime4}}\right)\nonumber \\
& \rightarrow \frac{\pi}{2}k^{\prime2}f(X^{2})X^{2}\left(1-X^{2}\right)^{2}k^{2}\hat{u}_{i}^{<}(\mathbf{q})-\frac{5}{2}\pi k^{\prime2}f(X^{2})X^{2}\left(1-X^{2}\right)^{2}P_{i3}k^{2}\hat{u}_{z}^{<}(\mathbf{q})\nonumber \\
 & \quad\;-\frac{5}{2}\pi k^{\prime2}f(X^{2})X^{2}\left(1-X^{2}\right)^{2}k_{z}^{2}\hat{u}_{i}^{<}(\mathbf{q})\nonumber\\
 & \quad\;-\frac{\pi}{2}k^{\prime2}f(X^{2})\left(1-X^{2}\right)\left(35X^{4}-35X^{2}+4\right)P_{i3}k_{z}^{2}\hat{u}_{z}^{<}(\mathbf{q}),\label{eq:aux4}
\end{align}
It is clear from the latter formulae, that a new term of the form
$P_{i3}k_{z}^{2}\hat{u}_{z}^{<}(\mathbf{q})$ appears in the equation
for the long-wavelength modes in the second step (\ref{eq:u<_second_step})
and hence we need to perform one more step, until invariance of the
equations for long-wavelength modes is achieved at each step and the
procedure can be closed and reduced to the form of recursion differential
equations. Hence we introduce $\lambda_{3}$, which satisfies
\begin{equation}
\delta\lambda=\lambda_{2}-\lambda_{3}\ll1,\label{eq:lambda_bite-1-1}
\end{equation}
and once again split the remaining fluctuational Fourier spectrum
$0\leq k\leq\lambda_{2}$ into two parts by defining new variables
(but keeping the same notation)
\begin{equation}
\theta(k-\lambda_{3})\hat{u}_{i}^{<}(\mathbf{k},\omega)\mapsto\hat{u}_{i}^{>}(\mathbf{k},\omega),\qquad(\textrm{for }\lambda_{3}<k<\lambda_{2}),\label{eq:Fourier_division_1-1-1}
\end{equation}
\begin{equation}
\theta(\lambda_{3}-k)\hat{u}_{i}^{<}(\mathbf{k},\omega)\mapsto\hat{u}_{i}^{<}(\mathbf{k},\omega),\qquad(\textrm{for }k<\lambda_{3}),\label{eq:Fourier_division_2-1-1}
\end{equation}
and same way for $\hat{Q}$. The equations are also split, as in the
first and second steps (cf. (\ref{eq:u<}) and (\ref{eq:u_gtr})),
i.e. for the new long-wavelength modes we have 
\begin{align}
& \left[-\mathrm{i}\omega+\breve{\xi}\left(\lambda_{2}\right)k^{2}+\breve{\zeta}\left(\lambda_{2}\right)k_{z}^{2}\right]\hat{u}_{i}^{<}(\mathbf{q})+\breve{\chi}\left(\lambda_{2}\right)P_{i3}k^{2}\hat{u}_{z}^{<}(\mathbf{q})+\breve{\varrho}\left(\lambda_{2}\right)P_{i3}k_{z}^{2}\hat{u}_{z}^{<}(\mathbf{q})=\qquad\qquad\qquad\qquad\qquad \nonumber\\
&\hspace{25mm} \frac{P_{i3}}{k^{2}}\hat{Q}^{<}(\mathbf{q})-\frac{1}{2}\mathrm{i}\epsilon P_{imn}\mathbb{I}_{mn}^{(u^{<})}\left(\mathbf{q}\right) -\frac{1}{2}\mathrm{i}\epsilon P_{imn}\int d^{4}q'\left\langle \hat{u}_{m}^{>}(\mathbf{q}')\hat{u}_{n}^{>}(\mathbf{q}-\mathbf{q}')\right\rangle _{c},\label{eq:u<_second_step-1}
\end{align}
Of course now $\left\langle \cdot\right\rangle _{c}$ denotes conditional
average over the second shell ($\lambda_{3}\leq k\leq\lambda_{2}$)
statistical subensemble but the coefficients $\breve{\xi}\left(\lambda_{2}\right)$,
$\breve{\zeta}\left(\lambda_{2}\right)$, $\breve{\chi}\left(\lambda_{2}\right)$
and $\breve{\varrho}\left(\lambda_{2}\right)$ are now strongly nonlinear
functions of $\breve{\xi}\left(\lambda_{1}\right)$, $\breve{\zeta}\left(\lambda_{1}\right)$,
$\breve{\chi}\left(\lambda_{1}\right)$. A third step is now necessary,
in order to verify, that in the following steps of the recursion procedure
the new term $\breve{\varrho}\left(\lambda_{2}\right)P_{i3}k_{z}^{2}\hat{u}_{z}^{<}(\mathbf{q})$
does not lead to appearance of yet other terms with distinct structure.
This however, is obvious since the new term can be treated as a correction
to the $\breve{\chi}\left(\lambda_{2}\right)$-term, thus for the
new short-wavelength modes we get (cf. (\ref{eq:u>_second_step}))
\begin{align}
\hat{u}_{i}^{>}(\mathbf{q})= & \frac{P_{i3}}{k^{2}\breve{\Gamma}}\hat{Q}^{>}(\mathbf{q})-\mathrm{i}\epsilon\frac{P_{imn}}{\breve{\gamma}}\mathbb{J}_{mn}^{(u)}\left(\mathbf{q}\right)+\mathrm{i}\epsilon\frac{\left(\breve{\chi}\left(\lambda_{1}\right)k^{2}+\breve{\varrho}\left(\lambda_{2}\right)k_{z}^{2}\right)}{\breve{\Gamma}}\frac{P_{i3}P_{3mn}}{\breve{\gamma}}\mathbb{J}_{mn}^{(u)}\left(\mathbf{q}\right)\nonumber \\
 & -\frac{1}{2}\mathrm{i}\epsilon\frac{P_{imn}}{\breve{\gamma}}\mathbb{I}_{mn}^{(u^{<})}\left(\mathbf{q}\right)+\frac{1}{2}\mathrm{i}\epsilon\frac{\left(\breve{\chi}\left(\lambda_{1}\right)k^{2}+\breve{\varrho}\left(\lambda_{2}\right)k_{z}^{2}\right)}{\breve{\Gamma}}\frac{P_{i3}P_{3mn}}{\breve{\gamma}}\mathbb{I}_{mn}^{(u^{<})}\left(\mathbf{q}\right)+R_{i},\label{eq:u>_second_step-1}
\end{align}
where
\begin{equation}
\mathbb{J}_{mn}^{(u)}\left(\mathbf{q}\right)=\int d^{4}q''\hat{u}_{m}^{<}(\mathbf{q}'')\hat{u}_{n}^{>}(\mathbf{q}-\mathbf{q}''),\label{eq:Ju-2-1}
\end{equation}
\begin{equation}
R_{i}=-\frac{1}{2}\mathrm{i}\epsilon\left[\frac{P_{imn}}{\breve{\gamma}}-\frac{\left(\breve{\chi}\left(\lambda_{1}\right)k^{2}+\breve{\varrho}\left(\lambda_{2}\right)k_{z}^{2}\right)}{\breve{\Gamma}}\frac{P_{i3}P_{3mn}}{\breve{\gamma}}\right]\mathbb{I}_{mn}^{(u^{>})}\left(\mathbf{q}\right).\label{eq:Ru-1-1}
\end{equation}
It follows, that although the new coefficient $\breve{\varrho}\left(\lambda_{2}\right)$
influences the dependencies of $\breve{\xi}\left(\lambda_{3}\right)$,
$\breve{\zeta}\left(\lambda_{3}\right)$, $\breve{\chi}\left(\lambda_{3}\right)$,
$\breve{\varrho}\left(\lambda_{3}\right)$ on $\breve{\xi}\left(\lambda_{2}\right)$,
$\breve{\zeta}\left(\lambda_{2}\right)$, $\breve{\chi}\left(\lambda_{2}\right)$
and $\breve{\varrho}\left(\lambda_{2}\right)$ and makes them even
more complex, no terms with new structure appear in the following
steps of the procedure. This allows to close the recursion problem
which results in some strongly nonlinear equations for the four coefficients
$\breve{\xi}\left(\lambda\right)$, $\breve{\zeta}\left(\lambda\right)$,
$\breve{\chi}\left(\lambda\right)$ and $\breve{\varrho}\left(\lambda\right)$
and hence the large-scale equations take the general form (\ref{eq:gen_form_MF}).

As a final note, it is to be stressed once again, that the entire technique fundamentally relies on two important assumptions, regarding the properties of the flow. Firstly we recall, that the statistical correlations between short-wavelength fluctuations of the order higher than second, i.e. terms of the type $\left\langle \hat{u}_{i}^{>}\hat{u}_{j}^{>}\hat{Q}^{>}\right\rangle _{c}$
have been neglected. Secondly, the limit of distant interactions (\ref{eq:dist_interact_limit})
corresponding to an assumption of ergodicity of the system has greatly simplified the calculations.


%
%

%


\bibliography{LSC_bibl}

\begin{thebibliography}{33}%
\makeatletter
\providecommand \@ifxundefined [1]{%
 \@ifx{#1\undefined}
}%
\providecommand \@ifnum [1]{%
 \ifnum #1\expandafter \@firstoftwo
 \else \expandafter \@secondoftwo
 \fi
}%
\providecommand \@ifx [1]{%
 \ifx #1\expandafter \@firstoftwo
 \else \expandafter \@secondoftwo
 \fi
}%
\providecommand \natexlab [1]{#1}%
\providecommand \enquote  [1]{``#1''}%
\providecommand \bibnamefont  [1]{#1}%
\providecommand \bibfnamefont [1]{#1}%
\providecommand \citenamefont [1]{#1}%
\providecommand \href@noop [0]{\@secondoftwo}%
\providecommand \href [0]{\begingroup \@sanitize@url \@href}%
\providecommand \@href[1]{\@@startlink{#1}\@@href}%
\providecommand \@@href[1]{\endgroup#1\@@endlink}%
\providecommand \@sanitize@url [0]{\catcode `\\12\catcode `\$12\catcode
  `\&12\catcode `\#12\catcode `\^12\catcode `\_12\catcode `\%12\relax}%
\providecommand \@@startlink[1]{}%
\providecommand \@@endlink[0]{}%
\providecommand \url  [0]{\begingroup\@sanitize@url \@url }%
\providecommand \@url [1]{\endgroup\@href {#1}{\urlprefix }}%
\providecommand \urlprefix  [0]{URL }%
\providecommand \Eprint [0]{\href }%
\providecommand \doibase [0]{http://dx.doi.org/}%
\providecommand \selectlanguage [0]{\@gobble}%
\providecommand \bibinfo  [0]{\@secondoftwo}%
\providecommand \bibfield  [0]{\@secondoftwo}%
\providecommand \translation [1]{[#1]}%
\providecommand \BibitemOpen [0]{}%
\providecommand \bibitemStop [0]{}%
\providecommand \bibitemNoStop [0]{.\EOS\space}%
\providecommand \EOS [0]{\spacefactor3000\relax}%
\providecommand \BibitemShut  [1]{\csname bibitem#1\endcsname}%
\let\auto@bib@innerbib\@empty
\bibitem [{\citenamefont {Zakharov}, \citenamefont {L{\textquoteright}vov},\
  and\ \citenamefont {Falkovich}(1992)}]{Zakharov_ea_1992}%
  \BibitemOpen
  \bibfield  {author} {\bibinfo {author} {\bibfnamefont {V.~E.}\ \bibnamefont
  {Zakharov}}, \bibinfo {author} {\bibfnamefont {V.~S.}\ \bibnamefont
  {L{\textquoteright}vov}}, \ and\ \bibinfo {author} {\bibfnamefont
  {G.}~\bibnamefont {Falkovich}},\ }\href@noop {} {\emph {\bibinfo {title}
  {Kolmogorov Spectra of Turbulence}}}\ (\bibinfo  {publisher} {Springer,
  Berlin},\ \bibinfo {year} {1992})\BibitemShut {NoStop}%
\bibitem [{\citenamefont {Newell}\ and\ \citenamefont
  {Rumpf}(2011)}]{NewellRumpf2011}%
  \BibitemOpen
  \bibfield  {author} {\bibinfo {author} {\bibfnamefont {A.~C.}\ \bibnamefont
  {Newell}}\ and\ \bibinfo {author} {\bibfnamefont {B.}~\bibnamefont {Rumpf}},\
  }\bibfield  {title} {\enquote {\bibinfo {title} {Wave turbulence},}\
  }\href@noop {} {\bibfield  {journal} {\bibinfo  {journal} {Annu. Rev. Fluid
  Mech.}\ }\textbf {\bibinfo {volume} {43}},\ \bibinfo {pages} {59--78}
  (\bibinfo {year} {2011})}\BibitemShut {NoStop}%
\bibitem [{\citenamefont {Nazarenko}(2011)}]{Nazarenko2011}%
  \BibitemOpen
  \bibfield  {author} {\bibinfo {author} {\bibfnamefont {S.}~\bibnamefont
  {Nazarenko}},\ }\href@noop {} {\emph {\bibinfo {title} {Wave Turbulence}}}\
  (\bibinfo  {publisher} {Springer, Berlin},\ \bibinfo {year}
  {2011})\BibitemShut {NoStop}%
\bibitem [{\citenamefont {Monsalve}\ \emph {et~al.}(2020)\citenamefont
  {Monsalve}, \citenamefont {Brunet}, \citenamefont {Gallet},\ and\
  \citenamefont {Cortet}}]{Monsalve_ea_2020}%
  \BibitemOpen
  \bibfield  {author} {\bibinfo {author} {\bibfnamefont {E.}~\bibnamefont
  {Monsalve}}, \bibinfo {author} {\bibfnamefont {M.}~\bibnamefont {Brunet}},
  \bibinfo {author} {\bibfnamefont {B.}~\bibnamefont {Gallet}}, \ and\ \bibinfo
  {author} {\bibfnamefont {P.-P.}\ \bibnamefont {Cortet}},\ }\bibfield  {title}
  {\enquote {\bibinfo {title} {Quantitative experimental observation of weak
  inertial-wave turbulence},}\ }\href@noop {} {\bibfield  {journal} {\bibinfo
  {journal} {Phys. Rev. Lett.}\ }\textbf {\bibinfo {volume} {125}},\ \bibinfo
  {pages} {254502} (\bibinfo {year} {2020})}\BibitemShut {NoStop}%
\bibitem [{\citenamefont {Tobias}, \citenamefont {Cattaneo},\ and\
  \citenamefont {Boldyrev}(2013)}]{Tobias_ea_2013}%
  \BibitemOpen
  \bibfield  {author} {\bibinfo {author} {\bibfnamefont {S.~M.}\ \bibnamefont
  {Tobias}}, \bibinfo {author} {\bibfnamefont {F.}~\bibnamefont {Cattaneo}}, \
  and\ \bibinfo {author} {\bibfnamefont {S.}~\bibnamefont {Boldyrev}},\
  }\href@noop {} {\emph {\bibinfo {title} {MHD dynamos and turbulence}}}\
  (\bibinfo  {publisher} {in Ten Chapters in Turbulence, P. A. Davidson, Y.
  Kaneda, and K. R. Sreenivasan, eds., Cambridge University Press, Cambridge,
  pp. 351--404},\ \bibinfo {year} {2013})\BibitemShut {NoStop}%
\bibitem [{\citenamefont {Wyld}(1961)}]{Wyld1961}%
  \BibitemOpen
  \bibfield  {author} {\bibinfo {author} {\bibfnamefont {J.~H.~W.}\
  \bibnamefont {Wyld}},\ }\bibfield  {title} {\enquote {\bibinfo {title}
  {Formulation of the theory of turbulence in an incompressible fluid},}\
  }\href@noop {} {\bibfield  {journal} {\bibinfo  {journal} {Annals of Phys.}\
  }\textbf {\bibinfo {volume} {14}},\ \bibinfo {pages} {143--165} (\bibinfo
  {year} {1961})}\BibitemShut {NoStop}%
\bibitem [{\citenamefont {Ma}\ and\ \citenamefont
  {Mazenko}(1975)}]{MaMazenko1975}%
  \BibitemOpen
  \bibfield  {author} {\bibinfo {author} {\bibfnamefont {S.}~\bibnamefont
  {Ma}}\ and\ \bibinfo {author} {\bibfnamefont {G.~F.}\ \bibnamefont
  {Mazenko}},\ }\bibfield  {title} {\enquote {\bibinfo {title} {Critical
  dynamics of ferromagnets in 6-$\epsilon$ dimensions: General discussion and
  detailed calculation},}\ }\href@noop {} {\bibfield  {journal} {\bibinfo
  {journal} {Phys. Rev. B}\ }\textbf {\bibinfo {volume} {11}},\ \bibinfo
  {pages} {4077--4100} (\bibinfo {year} {1975})}\BibitemShut {NoStop}%
\bibitem [{\citenamefont {Forster}, \citenamefont {Nelson},\ and\ \citenamefont
  {Stephen}(1977)}]{Forster_ea_1977}%
  \BibitemOpen
  \bibfield  {author} {\bibinfo {author} {\bibfnamefont {D.}~\bibnamefont
  {Forster}}, \bibinfo {author} {\bibfnamefont {D.~R.}\ \bibnamefont {Nelson}},
  \ and\ \bibinfo {author} {\bibfnamefont {M.~J.}\ \bibnamefont {Stephen}},\
  }\bibfield  {title} {\enquote {\bibinfo {title} {Large-distance and long-time
  properties of a randomly stirred fluid},}\ }\href@noop {} {\bibfield
  {journal} {\bibinfo  {journal} {Phys. Rev A}\ }\textbf {\bibinfo {volume}
  {16}},\ \bibinfo {pages} {732--749} (\bibinfo {year} {1977})}\BibitemShut
  {NoStop}%
\bibitem [{\citenamefont {Yakhot}\ and\ \citenamefont {Orszag}(1986)}]{YO1986}%
  \BibitemOpen
  \bibfield  {author} {\bibinfo {author} {\bibfnamefont {V.}~\bibnamefont
  {Yakhot}}\ and\ \bibinfo {author} {\bibfnamefont {S.~A.}\ \bibnamefont
  {Orszag}},\ }\bibfield  {title} {\enquote {\bibinfo {title} {Renormalization
  group analysis of turbulence. i. basic theory},}\ }\href@noop {} {\bibfield
  {journal} {\bibinfo  {journal} {J. Sci. Comp.}\ }\textbf {\bibinfo {volume}
  {1}},\ \bibinfo {pages} {3--51} (\bibinfo {year} {1986})}\BibitemShut
  {NoStop}%
\bibitem [{\citenamefont {Smith}\ and\ \citenamefont
  {Woodruff}(1998)}]{SW1998}%
  \BibitemOpen
  \bibfield  {author} {\bibinfo {author} {\bibfnamefont {L.~M.}\ \bibnamefont
  {Smith}}\ and\ \bibinfo {author} {\bibfnamefont {S.~L.}\ \bibnamefont
  {Woodruff}},\ }\bibfield  {title} {\enquote {\bibinfo {title}
  {Renormalization-group analysis of turbulence},}\ }\href@noop {} {\bibfield
  {journal} {\bibinfo  {journal} {Annu. Rev. Fluid Mech.}\ }\textbf {\bibinfo
  {volume} {30}},\ \bibinfo {pages} {275--310} (\bibinfo {year}
  {1998})}\BibitemShut {NoStop}%
\bibitem [{\citenamefont {McComb}(2014)}]{McComb2014}%
  \BibitemOpen
  \bibfield  {author} {\bibinfo {author} {\bibfnamefont {W.~D.}\ \bibnamefont
  {McComb}},\ }\href@noop {} {\emph {\bibinfo {title} {Homogeneous, isotropic
  turbulence. Phenomenology, renormalization and statistical closures}}}\
  (\bibinfo  {publisher} {Oxford University Press, Oxford},\ \bibinfo {year}
  {2014})\BibitemShut {NoStop}%
\bibitem [{\citenamefont {Kraichnan}(1959)}]{Kraichnan1959}%
  \BibitemOpen
  \bibfield  {author} {\bibinfo {author} {\bibfnamefont {R.~H.}\ \bibnamefont
  {Kraichnan}},\ }\bibfield  {title} {\enquote {\bibinfo {title} {The structure
  of isotropic turbulence at very high reynolds numbers},}\ }\href@noop {}
  {\bibfield  {journal} {\bibinfo  {journal} {J. Fluid Mech}\ }\textbf
  {\bibinfo {volume} {5}},\ \bibinfo {pages} {497--543} (\bibinfo {year}
  {1959})}\BibitemShut {NoStop}%
\bibitem [{\citenamefont {Kraichnan}(1965)}]{Kraichnan1965}%
  \BibitemOpen
  \bibfield  {author} {\bibinfo {author} {\bibfnamefont {R.~H.}\ \bibnamefont
  {Kraichnan}},\ }\bibfield  {title} {\enquote {\bibinfo {title}
  {Lagrangian-history closure approximation for turbulence},}\ }\href@noop {}
  {\bibfield  {journal} {\bibinfo  {journal} {Phys. Fluids}\ }\textbf {\bibinfo
  {volume} {8}},\ \bibinfo {pages} {575--598} (\bibinfo {year}
  {1965})}\BibitemShut {NoStop}%
\bibitem [{\citenamefont {Yoshizawa}(1998)}]{Yoshizawa1998}%
  \BibitemOpen
  \bibfield  {author} {\bibinfo {author} {\bibfnamefont {A.}~\bibnamefont
  {Yoshizawa}},\ }\href@noop {} {\emph {\bibinfo {title} {Hydrodynamic and
  Magnetohydrodynamic Turbulent Flows: Modelling and Statistical Theory}}}\
  (\bibinfo  {publisher} {Kluwer Academic Publishers},\ \bibinfo {year}
  {1998})\BibitemShut {NoStop}%
\bibitem [{\citenamefont {Yokoi}(2020)}]{Yokoi2020}%
  \BibitemOpen
  \bibfield  {author} {\bibinfo {author} {\bibfnamefont {N.}~\bibnamefont
  {Yokoi}},\ }\href@noop {} {\emph {\bibinfo {title} {Turbulence, transport and
  reconnection}}}\ (\bibinfo  {publisher} {in Topics in Magnetohydrodynamic
  Topology, Reconnection and Stability Theory (ed. D. MacTaggart \& A.
  Hillier), vol. 59. CISM International Centre for Mechanical Sciences,
  Springer},\ \bibinfo {year} {2020})\BibitemShut {NoStop}%
\bibitem [{\citenamefont {Mizerski}(2020)}]{Mizerski2020}%
  \BibitemOpen
  \bibfield  {author} {\bibinfo {author} {\bibfnamefont {K.~A.}\ \bibnamefont
  {Mizerski}},\ }\bibfield  {title} {\enquote {\bibinfo {title}
  {Renormalization group analysis of the turbulent hydromagnetic dynamo: effect
  of non-stationarity},}\ }\href@noop {} {\bibfield  {journal} {\bibinfo
  {journal} {Astroph. J. Suppl. Ser.}\ }\textbf {\bibinfo {volume} {251}},\
  \bibinfo {pages} {21 (29pp)} (\bibinfo {year} {2020})}\BibitemShut {NoStop}%
\bibitem [{\citenamefont {Mizerski}(021a)}]{Mizerski2021a}%
  \BibitemOpen
  \bibfield  {author} {\bibinfo {author} {\bibfnamefont {K.~A.}\ \bibnamefont
  {Mizerski}},\ }\bibfield  {title} {\enquote {\bibinfo {title}
  {Renormalization group analysis of the turbulent hydromagnetic dynamo: effect
  of anizotropy},}\ }\href@noop {} {\bibfield  {journal} {\bibinfo  {journal}
  {Appl. Math. Comput.}\ }\textbf {\bibinfo {volume} {405}},\ \bibinfo {pages}
  {126252} (\bibinfo {year} {2021a})}\BibitemShut {NoStop}%
\bibitem [{\citenamefont {Camps}\ \emph {et~al.}(2015)\citenamefont {Camps},
  \citenamefont {Misselt}, \citenamefont {Bianchi}, \citenamefont {Lunttila},
  \citenamefont {Pinte}, \citenamefont {Natale}, \citenamefont {Juvela},
  \citenamefont {Fischera}, \citenamefont {Fitzgerald}, \citenamefont {Gordon},
  \citenamefont {Baes},\ and\ \citenamefont {Steinacker}}]{Camps_ea_2015}%
  \BibitemOpen
  \bibfield  {author} {\bibinfo {author} {\bibfnamefont {P.}~\bibnamefont
  {Camps}}, \bibinfo {author} {\bibfnamefont {K.}~\bibnamefont {Misselt}},
  \bibinfo {author} {\bibfnamefont {S.}~\bibnamefont {Bianchi}}, \bibinfo
  {author} {\bibfnamefont {T.}~\bibnamefont {Lunttila}}, \bibinfo {author}
  {\bibfnamefont {C.}~\bibnamefont {Pinte}}, \bibinfo {author} {\bibfnamefont
  {G.}~\bibnamefont {Natale}}, \bibinfo {author} {\bibfnamefont
  {M.}~\bibnamefont {Juvela}}, \bibinfo {author} {\bibfnamefont
  {J.}~\bibnamefont {Fischera}}, \bibinfo {author} {\bibfnamefont {M.~P.}\
  \bibnamefont {Fitzgerald}}, \bibinfo {author} {\bibfnamefont
  {K.}~\bibnamefont {Gordon}}, \bibinfo {author} {\bibfnamefont
  {M.}~\bibnamefont {Baes}}, \ and\ \bibinfo {author} {\bibfnamefont
  {J.}~\bibnamefont {Steinacker}},\ }\bibfield  {title} {\enquote {\bibinfo
  {title} {Benchmarking the calculation of stochastic heating and emissivity of
  dust grains in the context of radiative transfer simulations},}\ }\href@noop
  {} {\bibfield  {journal} {\bibinfo  {journal} {Astron. Astrophys.}\ }\textbf
  {\bibinfo {volume} {580}},\ \bibinfo {pages} {A87} (\bibinfo {year}
  {2015})}\BibitemShut {NoStop}%
\bibitem [{\citenamefont {Serra-Garcia}\ \emph {et~al.}(2016)\citenamefont
  {Serra-Garcia}, \citenamefont {Foehr}, \citenamefont
  {Moler$\acute{\textrm{o}}$n}, \citenamefont {Lydon}, \citenamefont {Chong},\
  and\ \citenamefont {Daraio}}]{Serra-Garcia_ea_2016}%
  \BibitemOpen
  \bibfield  {author} {\bibinfo {author} {\bibfnamefont {M.}~\bibnamefont
  {Serra-Garcia}}, \bibinfo {author} {\bibfnamefont {A.}~\bibnamefont {Foehr}},
  \bibinfo {author} {\bibfnamefont {M.}~\bibnamefont
  {Moler$\acute{\textrm{o}}$n}}, \bibinfo {author} {\bibfnamefont
  {J.}~\bibnamefont {Lydon}}, \bibinfo {author} {\bibfnamefont
  {C.}~\bibnamefont {Chong}}, \ and\ \bibinfo {author} {\bibfnamefont
  {C.}~\bibnamefont {Daraio}},\ }\bibfield  {title} {\enquote {\bibinfo {title}
  {Mechanical autonomous stochastic heat engine},}\ }\href@noop {} {\bibfield
  {journal} {\bibinfo  {journal} {Phys. Rev. Lett.}\ }\textbf {\bibinfo
  {volume} {117}},\ \bibinfo {pages} {010602} (\bibinfo {year}
  {2016})}\BibitemShut {NoStop}%
\bibitem [{\citenamefont {Grossmann}\ and\ \citenamefont
  {Lohse}(2000)}]{GrossmannLohse2000}%
  \BibitemOpen
  \bibfield  {author} {\bibinfo {author} {\bibfnamefont {S.}~\bibnamefont
  {Grossmann}}\ and\ \bibinfo {author} {\bibfnamefont {D.}~\bibnamefont
  {Lohse}},\ }\bibfield  {title} {\enquote {\bibinfo {title} {Scaling in
  thermal convection: a unifying theory},}\ }\href@noop {} {\bibfield
  {journal} {\bibinfo  {journal} {J. Fluid Mech.}\ }\textbf {\bibinfo {volume}
  {407}},\ \bibinfo {pages} {27--56} (\bibinfo {year} {2000})}\BibitemShut
  {NoStop}%
\bibitem [{\citenamefont {Funfschilling}\ and\ \citenamefont
  {Ahlers}(2004)}]{FunfschillingAhlers2004}%
  \BibitemOpen
  \bibfield  {author} {\bibinfo {author} {\bibfnamefont {D.}~\bibnamefont
  {Funfschilling}}\ and\ \bibinfo {author} {\bibfnamefont {G.}~\bibnamefont
  {Ahlers}},\ }\bibfield  {title} {\enquote {\bibinfo {title} {Plume motion and
  large-scale circulation in a cylindrical rayleigh-b$\acute{\textrm{e}}$nard
  cell},}\ }\href@noop {} {\bibfield  {journal} {\bibinfo  {journal} {Phys. Rev
  Lett.}\ }\textbf {\bibinfo {volume} {92}},\ \bibinfo {pages} {194502}
  (\bibinfo {year} {2004})}\BibitemShut {NoStop}%
\bibitem [{\citenamefont {Brown}\ and\ \citenamefont
  {Ahlers}(2009)}]{BrownAhlers2009}%
  \BibitemOpen
  \bibfield  {author} {\bibinfo {author} {\bibfnamefont {E.}~\bibnamefont
  {Brown}}\ and\ \bibinfo {author} {\bibfnamefont {G.}~\bibnamefont {Ahlers}},\
  }\bibfield  {title} {\enquote {\bibinfo {title} {The origin of oscillations
  of the large-scale circulation of turbulent rayleigh\textendash
  b$\acute{\textrm{e}}$nard convection},}\ }\href@noop {} {\bibfield  {journal}
  {\bibinfo  {journal} {J. Fluid Mech.}\ }\textbf {\bibinfo {volume} {638}},\
  \bibinfo {pages} {383--400} (\bibinfo {year} {2009})}\BibitemShut {NoStop}%
\bibitem [{\citenamefont {Xi}\ \emph {et~al.}(2009)\citenamefont {Xi},
  \citenamefont {Zhou}, \citenamefont {Zhou}, \citenamefont {Chan},\ and\
  \citenamefont {Xia}}]{Xi_ea_2009}%
  \BibitemOpen
  \bibfield  {author} {\bibinfo {author} {\bibfnamefont {H.-D.}\ \bibnamefont
  {Xi}}, \bibinfo {author} {\bibfnamefont {S.-Q.}\ \bibnamefont {Zhou}},
  \bibinfo {author} {\bibfnamefont {Q.}~\bibnamefont {Zhou}}, \bibinfo {author}
  {\bibfnamefont {T.-S.}\ \bibnamefont {Chan}}, \ and\ \bibinfo {author}
  {\bibfnamefont {K.-Q.}\ \bibnamefont {Xia}},\ }\bibfield  {title} {\enquote
  {\bibinfo {title} {Origin of the temperature oscillation in turbulent thermal
  convection},}\ }\href@noop {} {\bibfield  {journal} {\bibinfo  {journal}
  {Phys. Rev. Lett.}\ }\textbf {\bibinfo {volume} {102}},\ \bibinfo {pages}
  {044503} (\bibinfo {year} {2009})}\BibitemShut {NoStop}%
\bibitem [{\citenamefont {Horn}, \citenamefont {Schmid},\ and\ \citenamefont
  {Aurnou}(2022)}]{Horn_ea_2022}%
  \BibitemOpen
  \bibfield  {author} {\bibinfo {author} {\bibfnamefont {S.}~\bibnamefont
  {Horn}}, \bibinfo {author} {\bibfnamefont {P.~J.}\ \bibnamefont {Schmid}}, \
  and\ \bibinfo {author} {\bibfnamefont {J.~M.}\ \bibnamefont {Aurnou}},\
  }\bibfield  {title} {\enquote {\bibinfo {title} {Unravelling the large-scale
  circulation modes in turbulent rayleigh-b$\acute{\textrm{e}}$nard
  convection},}\ }\href@noop {} {\bibfield  {journal} {\bibinfo  {journal}
  {Europhys. Lett.}\ }\textbf {\bibinfo {volume} {136}},\ \bibinfo {pages}
  {14003} (\bibinfo {year} {2022})}\BibitemShut {NoStop}%
\bibitem [{\citenamefont {Roche}(2020)}]{Roche2020}%
  \BibitemOpen
  \bibfield  {author} {\bibinfo {author} {\bibfnamefont {P.-E.}\ \bibnamefont
  {Roche}},\ }\bibfield  {title} {\enquote {\bibinfo {title} {The ultimate
  state of convection: a unifying picture of very high rayleigh numbers
  experiments},}\ }\href@noop {} {\bibfield  {journal} {\bibinfo  {journal}
  {New J. Phys.}\ }\textbf {\bibinfo {volume} {22}},\ \bibinfo {pages} {073056}
  (\bibinfo {year} {2020})}\BibitemShut {NoStop}%
\bibitem [{\citenamefont {Zwirner}, \citenamefont {Tilgner},\ and\
  \citenamefont {Shishkina}(2020)}]{Zwirner_ea_2020}%
  \BibitemOpen
  \bibfield  {author} {\bibinfo {author} {\bibfnamefont {L.}~\bibnamefont
  {Zwirner}}, \bibinfo {author} {\bibfnamefont {A.}~\bibnamefont {Tilgner}}, \
  and\ \bibinfo {author} {\bibfnamefont {O.}~\bibnamefont {Shishkina}},\
  }\bibfield  {title} {\enquote {\bibinfo {title} {Elliptical instability and
  multiple-roll flow modes of the large-scale circulation in confined turbulent
  rayleigh-b$\acute{\textrm{e}}$nard convection},}\ }\href@noop {} {\bibfield
  {journal} {\bibinfo  {journal} {Phys. Rev. Lett.}\ }\textbf {\bibinfo
  {volume} {125}},\ \bibinfo {pages} {054502} (\bibinfo {year}
  {2020})}\BibitemShut {NoStop}%
\bibitem [{\citenamefont {Vasiliev}\ \emph {et~al.}(2019)\citenamefont
  {Vasiliev}, \citenamefont {Frick}, \citenamefont {Kumar}, \citenamefont
  {Stepanov}, \citenamefont {Sukhanovskii},\ and\ \citenamefont
  {Verma}}]{Vasiliev_ea_2019}%
  \BibitemOpen
  \bibfield  {author} {\bibinfo {author} {\bibfnamefont {A.}~\bibnamefont
  {Vasiliev}}, \bibinfo {author} {\bibfnamefont {P.}~\bibnamefont {Frick}},
  \bibinfo {author} {\bibfnamefont {A.}~\bibnamefont {Kumar}}, \bibinfo
  {author} {\bibfnamefont {R.}~\bibnamefont {Stepanov}}, \bibinfo {author}
  {\bibfnamefont {A.}~\bibnamefont {Sukhanovskii}}, \ and\ \bibinfo {author}
  {\bibfnamefont {M.~K.}\ \bibnamefont {Verma}},\ }\bibfield  {title} {\enquote
  {\bibinfo {title} {Transient flows and reorientations of large-scale
  convection in a cubic cell},}\ }\href@noop {} {\bibfield  {journal} {\bibinfo
   {journal} {Int. Comm. Heat Mass Trans.}\ }\textbf {\bibinfo {volume}
  {108}},\ \bibinfo {pages} {104319} (\bibinfo {year} {2019})}\BibitemShut
  {NoStop}%
\bibitem [{\citenamefont {Spiegel}\ and\ \citenamefont
  {Veronis}(1960)}]{SpiegelVeronis1960}%
  \BibitemOpen
  \bibfield  {author} {\bibinfo {author} {\bibfnamefont {E.~A.}\ \bibnamefont
  {Spiegel}}\ and\ \bibinfo {author} {\bibfnamefont {G.}~\bibnamefont
  {Veronis}},\ }\bibfield  {title} {\enquote {\bibinfo {title} {On the
  boussinesq approximation for a compressible fluid},}\ }\href@noop {}
  {\bibfield  {journal} {\bibinfo  {journal} {Astrophys. J.}\ }\textbf
  {\bibinfo {volume} {131}},\ \bibinfo {pages} {442--447} (\bibinfo {year}
  {1960})}\BibitemShut {NoStop}%
\bibitem [{\citenamefont {Mizerski}(021b)}]{Mizerski2021b}%
  \BibitemOpen
  \bibfield  {author} {\bibinfo {author} {\bibfnamefont {K.~A.}\ \bibnamefont
  {Mizerski}},\ }\href@noop {} {\emph {\bibinfo {title} {Foundations of
  convection with density stratification}}}\ (\bibinfo  {publisher} {Springer
  Nature, Cham, Switzerland},\ \bibinfo {year} {2021b})\BibitemShut {NoStop}%
\bibitem [{\citenamefont {McComb}, \citenamefont {Roberts},\ and\ \citenamefont
  {Watt}(1992)}]{McComb_ea_1992}%
  \BibitemOpen
  \bibfield  {author} {\bibinfo {author} {\bibfnamefont {W.~D.}\ \bibnamefont
  {McComb}}, \bibinfo {author} {\bibfnamefont {W.}~\bibnamefont {Roberts}}, \
  and\ \bibinfo {author} {\bibfnamefont {A.~G.}\ \bibnamefont {Watt}},\
  }\bibfield  {title} {\enquote {\bibinfo {title} {Conditional-averaging
  procedure for problems with mode-mode coupling},}\ }\href@noop {} {\bibfield
  {journal} {\bibinfo  {journal} {Phys. Rev. A}\ }\textbf {\bibinfo {volume}
  {45}},\ \bibinfo {pages} {3507--3515} (\bibinfo {year} {1992})}\BibitemShut
  {NoStop}%
\bibitem [{\citenamefont {McComb}\ and\ \citenamefont
  {Watt}(1990)}]{McCombWatt1990}%
  \BibitemOpen
  \bibfield  {author} {\bibinfo {author} {\bibfnamefont {W.~D.}\ \bibnamefont
  {McComb}}\ and\ \bibinfo {author} {\bibfnamefont {A.~G.}\ \bibnamefont
  {Watt}},\ }\bibfield  {title} {\enquote {\bibinfo {title} {Conditional
  averaging procedure for the elimination of the small-scale modes from
  incompressible fluid turbulence at high reynolds numbers},}\ }\href@noop {}
  {\bibfield  {journal} {\bibinfo  {journal} {Phys. Rev. Lett.}\ }\textbf
  {\bibinfo {volume} {65}},\ \bibinfo {pages} {3281--3284} (\bibinfo {year}
  {1990})}\BibitemShut {NoStop}%
\bibitem [{\citenamefont {McComb}\ and\ \citenamefont
  {Watt}(1992)}]{McCombWatt1992}%
  \BibitemOpen
  \bibfield  {author} {\bibinfo {author} {\bibfnamefont {W.~D.}\ \bibnamefont
  {McComb}}\ and\ \bibinfo {author} {\bibfnamefont {A.~G.}\ \bibnamefont
  {Watt}},\ }\bibfield  {title} {\enquote {\bibinfo {title} {Two-field theory
  of incompressible-fluid turbulence},}\ }\href@noop {} {\bibfield  {journal}
  {\bibinfo  {journal} {Phys. Rev. A}\ }\textbf {\bibinfo {volume} {46}},\
  \bibinfo {pages} {4797--4812} (\bibinfo {year} {1992})}\BibitemShut {NoStop}%
\bibitem [{\citenamefont {Lifshitz}\ and\ \citenamefont
  {Pitaevskii}(1987)}]{LifshitzPitaevskii1987}%
  \BibitemOpen
  \bibfield  {author} {\bibinfo {author} {\bibfnamefont {E.~M.}\ \bibnamefont
  {Lifshitz}}\ and\ \bibinfo {author} {\bibfnamefont {L.~P.}\ \bibnamefont
  {Pitaevskii}},\ }\href@noop {} {\emph {\bibinfo {title} {Statistical physics,
  part 2, Landau and Lifshitz course of theoretical physics, vol. 9}}}\
  (\bibinfo  {publisher} {Butterworth-Heinemann, Oxford},\ \bibinfo {year}
  {1987})\BibitemShut {NoStop}%
\end{thebibliography}%

\end{document}